\documentclass[trackchanges, twocolumn]{aastex701}

\usepackage{tabularx}
\usepackage{tablefootnote}
\usepackage{CJKutf8}
\usepackage{float}

\begin{document}

\title{The Eye of Sauron in SN\,2025ngs: a Short-plateau Cousin of SN\,1998S with Evidence for a Ring-like Circumstellar Medium}

\newcommand{\LCO}{\affiliation{Las Cumbres Observatory, 6740 Cortona Drive, Suite 102, Goleta, CA 93117-5575, USA}}
\newcommand{\UCSB}{\affiliation{Department of Physics, University of California, Santa Barbara, CA 93106-9530, USA}}
\newcommand{\UCSD}{\affiliation{Department of Astronomy \& Astrophysics, University of California, San Diego, 9500 Gilman Drive, MC 0424, La Jolla, CA 92093-0424, USA}}
\newcommand{\KITP}{\affiliation{Kavli Institute for Theoretical Physics, University of California, Santa Barbara, CA 93106-4030, USA}}
\newcommand{\UCD}{\affiliation{Department of Physics and Astronomy, University of California, Davis, 1 Shields Avenue, Davis, CA 95616-5270, USA}}
\newcommand{\WIS}{\affiliation{Department of Particle Physics and Astrophysics, Weizmann Institute of Science, 76100 Rehovot, Israel}}
\newcommand{\OKC}{\affiliation{Oskar Klein Centre, Department of Astronomy, Stockholm University, Albanova University Centre, SE-106 91 Stockholm, Sweden}}
\newcommand{\OAPD}{\affiliation{INAF-Osservatorio Astronomico di Padova, Vicolo dell'Osservatorio 5, I-35122 Padova, Italy}}
\newcommand{\OAB}{\affiliation{INAF-Osservatorio Astronomico di Brera, Via E. Bianchi 46, I-23807, Merate (LC), Italy}}
\newcommand{\Caltech}{\affiliation{Cahill Center for Astronomy and Astrophysics, California Institute of Technology, Mail Code 249-17, Pasadena, CA 91125, USA}}
\newcommand{\GSFC}{\affiliation{Astrophysics Science Division, NASA Goddard Space Flight Center, Mail Code 661, Greenbelt, MD 20771, USA}}
\newcommand{\UMD}{\affiliation{Joint Space-Science Institute, University of Maryland, College Park, MD 20742, USA}}
\newcommand{\UCB}{\affiliation{Department of Astronomy, University of California, Berkeley, CA 94720-3411, USA}}
\newcommand{\TTU}{\affiliation{Department of Physics, Texas Tech University, Box 41051, Lubbock, TX 79409-1051, USA}}
\newcommand{\STScI}{\affiliation{Space Telescope Science Institute, 3700 San Martin Drive, Baltimore, MD 21218-2410, USA}}
\newcommand{\UT}{\affiliation{University of Texas at Austin, 1 University Station C1400, Austin, TX 78712-0259, USA}}
\newcommand{\IoA}{\affiliation{Institute of Astronomy, University of Cambridge, Madingley Road, Cambridge CB3 0HA, UK}}
\newcommand{\QUB}{\affiliation{Astrophysics Research Centre, School of Mathematics and Physics, Queen's University Belfast, Belfast BT7 1NN, UK}}
\newcommand{\IPAC}{\affiliation{IPAC, Mail Code 100-22, Caltech, 1200 E.\ California Blvd., Pasadena, CA 91125}}
\newcommand{\JPL}{\affiliation{Jet Propulsion Laboratory, California Institute of Technology, 4800 Oak Grove Dr, Pasadena, CA 91109, USA}}
\newcommand{\Southampton}{\affiliation{Department of Physics and Astronomy, University of Southampton, Southampton SO17 1BJ, UK}}
\newcommand{\LANL}{\affiliation{Space and Remote Sensing, MS B244, Los Alamos National Laboratory, Los Alamos, NM 87545, USA}}
\newcommand{\Tsinghua}{\affiliation{Physics Department and Tsinghua Center for Astrophysics, Tsinghua University, Beijing, 100084, People's Republic of China}}
\newcommand{\NAOC}{\affiliation{National Astronomical Observatory of China, Chinese Academy of Sciences, Beijing, 100012, People's Republic of China}}
\newcommand{\Itagaki}{\affiliation{Itagaki Astronomical Observatory, Yamagata 990-2492, Japan}}
\newcommand{\Einstein}{\altaffiliation{Einstein Fellow}}
\newcommand{\Hubble}{\altaffiliation{Hubble Fellow}}
\newcommand{\CfA}{\affiliation{Center for Astrophysics \textbar{} Harvard \& Smithsonian, 60 Garden Street, Cambridge, MA 02138-1516, USA}}
\newcommand{\UA}{\affiliation{Steward Observatory, University of Arizona, 933 North Cherry Avenue, Tucson, AZ 85721-0065, USA}}
\newcommand{\MPIA}{\affiliation{Max-Planck-Institut f\"ur Astrophysik, Karl-Schwarzschild-Stra\ss{}e 1, D-85748 Garching, Germany}}
\newcommand{\DSFP}{\altaffiliation{LSSTC Data Science Fellow}}
\newcommand{\HCO}{\affiliation{Harvard College Observatory, 60 Garden Street, Cambridge, MA 02138-1516, USA}}
\newcommand{\Carnegie}{\affiliation{Observatories of the Carnegie Institute for Science, 813 Santa Barbara Street, Pasadena, CA 91101-1232, USA}}
\newcommand{\TAU}{\affiliation{School of Physics and Astronomy, Tel Aviv University, Tel Aviv 69978, Israel}}
\newcommand{\Edinburgh}{\affiliation{Institute for Astronomy, University of Edinburgh, Royal Observatory, Blackford Hill EH9 3HJ, UK}}
\newcommand{\Birmingham}{\affiliation{Birmingham Institute for Gravitational Wave Astronomy and School of Physics and Astronomy, University of Birmingham, Birmingham B15 2TT, UK}}
\newcommand{\Bath}{\affiliation{Department of Physics, University of Bath, Claverton Down, Bath BA2 7AY, UK}}
\newcommand{\CTIO}{\affiliation{Cerro Tololo Inter-American Observatory, National Optical Astronomy Observatory, Casilla 603, La Serena, Chile}}
\newcommand{\Potsdam}{\affiliation{Leibniz-Institut f\"ur Astrophysik Potsdam (AIP), An der Sternwarte 16, D-14482 Potsdam, Germany}}
\newcommand{\INPE}{\affiliation{Instituto Nacional de Pesquisas Espaciais, Avenida dos Astronautas 1758, 12227-010, S\~ao Jos\'e dos Campos -- SP, Brazil}}
\newcommand{\UNC}{\affiliation{Department of Physics and Astronomy, University of North Carolina, 120 East Cameron Avenue, Chapel Hill, NC 27599, USA}}
\newcommand{\Ohio}{\affiliation{Astrophysical Institute, Department of Physics and Astronomy, 251B Clippinger Lab, Ohio University, Athens, OH 45701-2942, USA}}
\newcommand{\AAS}{\affiliation{American Astronomical Society, 1667 K~Street NW, Suite 800, Washington, DC 20006-1681, USA}}
\newcommand{\MMT}{\affiliation{MMT and Steward Observatories, University of Arizona, 933 North Cherry Avenue, Tucson, AZ 85721-0065, USA}}
\newcommand{\Geneva}{\affiliation{ISDC, Department of Astronomy, University of Geneva, Chemin d'\'Ecogia, 16 CH-1290 Versoix, Switzerland}}
\newcommand{\IUCAA}{\affiliation{Inter-University Center for Astronomy and Astrophysics, Post Bag 4, Ganeshkhind, Pune, Maharashtra 411007, India}}
\newcommand{\CMU}{\affiliation{Department of Physics, Carnegie Mellon University, 5000 Forbes Avenue, Pittsburgh, PA 15213-3815, USA}}
\newcommand{\NAOJ}{\affiliation{Division of Science, National Astronomical Observatory of Japan, 2-21-1 Osawa, Mitaka, Tokyo 181-8588, Japan}}
\newcommand{\IfA}{\affiliation{Institute for Astronomy, University of Hawai`i, 2680 Woodlawn Drive, Honolulu, HI 96822-1839, USA}}
\newcommand{\UCSC}{\affiliation{Department of Astronomy and Astrophysics, University of California, Santa Cruz, CA 95064-1077, USA}}
\newcommand{\Purdue}{\affiliation{Department of Physics and Astronomy, Purdue University, 525 Northwestern Avenue, West Lafayette, IN 47907-2036, USA}}
\newcommand{\Princeton}{\affiliation{Department of Astrophysical Sciences, Princeton University, 4 Ivy Lane, Princeton, NJ 08540-7219, USA}}
\newcommand{\Moore}{\affiliation{Gordon and Betty Moore Foundation, 1661 Page Mill Road, Palo Alto, CA 94304-1209, USA}}
\newcommand{\Durham}{\affiliation{Department of Physics, Durham University, South Road, Durham, DH1 3LE, UK}}
\newcommand{\JHU}{\affiliation{Department of Physics and Astronomy, The Johns Hopkins University, 3400 North Charles Street, Baltimore, MD 21218, USA}}
\newcommand{\Toronto}{\affiliation{David A.\ Dunlap Department of Astronomy and Astrophysics, University of Toronto,\\ 50 St.\ George Street, Toronto, Ontario, M5S 3H4 Canada}}
\newcommand{\Duke}{\affiliation{Department of Physics, Duke University, Campus Box 90305, Durham, NC 27708, USA}}
\newcommand{\NCU}{\affiliation{Graduate Institute of Astronomy, National Central University, 300 Jhongda Road, 32001 Jhongli, Taiwan}}
\newcommand{\Columbia}{\affiliation{Department of Physics and Columbia Astrophysics Laboratory, Columbia University, Pupin Hall, New York, NY 10027, USA}}
\newcommand{\Flatiron}{\affiliation{Center for Computational Astrophysics, Flatiron Institute, 162 5th Avenue, New York, NY 10010-5902, USA}}
\newcommand{\NU}{\affiliation{Department of Physics and Astronomy, Northwestern University, 2145 Sheridan Rd, Evanston, IL 60208, USA}}
\newcommand{\CIERA}{\affiliation{Center for Interdisciplinary Exploration and Research in Astrophysics, \\Northwestern University, 1800 Sherman Avenue, 8th Floor, Evanston, IL 60201, USA}}
\newcommand{\SkAI}{\affil{NSF-Simons AI Institute for the Sky (SkAI), 172 E. Chestnut St., Chicago, IL 60611, USA}}
\newcommand{\GeminiNorth}{\affiliation{Gemini Observatory, 670 North A`ohoku Place, Hilo, HI 96720-2700, USA}}
\newcommand{\Keck}{\affiliation{W.~M.~Keck Observatory, 65-1120 M\=amalahoa Highway, Kamuela, HI 96743-8431, USA}}
\newcommand{\UW}{\affiliation{Department of Astronomy, University of Washington, 3910 15th Avenue NE, Seattle, WA 98195-0002, USA}}
\newcommand{\Catalyst}{\altaffiliation{LSST-DA Catalyst Fellow}}
\newcommand{\USask}{\affiliation{Department of Physics and Engineering Physics, University of Saskatchewan, 116 Science Place, Saskatoon, SK S7N 5E2, Canada}}
\newcommand{\Thacher}{\affiliation{Thacher School, 5025 Thacher Road, Ojai, CA 93023-8304, USA}}
\newcommand{\Rutgers}{\affiliation{Department of Physics and Astronomy, Rutgers, the State University of New Jersey,\\136 Frelinghuysen Road, Piscataway, NJ 08854-8019, USA}}
\newcommand{\FSU}{\affiliation{Department of Physics, Florida State University, 77 Chieftan Way, Tallahassee, FL 32306-4350, USA}}
\newcommand{\Melbourne}{\affiliation{School of Physics, The University of Melbourne, Parkville, VIC 3010, Australia}}
\newcommand{\ASTROthreeD}{\affiliation{ARC Centre of Excellence for All Sky Astrophysics in 3 Dimensions (ASTRO 3D)}}
\newcommand{\Stromlo}{\affiliation{Mt.\ Stromlo Observatory, The Research School of Astronomy and Astrophysics, Australian National University, ACT 2601, Australia}}
\newcommand{\NCPAS}{\affiliation{National Centre for the Public Awareness of Science, Australian National University, Canberra, ACT 2611, Australia}}
\newcommand{\TAMU}{\affiliation{Department of Physics and Astronomy, Texas A\&M University, 4242 TAMU, College Station, TX 77843, USA}}
\newcommand{\Mitchell}{\affiliation{George P.\ and Cynthia Woods Mitchell Institute for Fundamental Physics \& Astronomy, College Station, TX 77843, USA}}
\newcommand{\ESO}{\affiliation{European Southern Observatory, Alonso de C\'ordova 3107, Casilla 19, Santiago, Chile}}
\newcommand{\ICE}{\affiliation{Institute of Space Sciences (ICE, CSIC), Campus UAB, Carrer
de Can Magrans, s/n, E-08193 Barcelona, Spain}}
\newcommand{\IEEC}{\affiliation{Institut d'Estudis Espacials de Catalunya, Gran Capit\`a, 2-4, Edifici Nexus, Desp.\ 201, E-08034 Barcelona, Spain}}
\newcommand{\Warwick}{\affiliation{Department of Physics, University of Warwick, Gibbet Hill Road, Coventry CV4 7AL, UK}}
\newcommand{\Macquarie}{\affiliation{School of Mathematical and Physical Sciences, Macquarie University, NSW 2109, Australia}}
\newcommand{\AAARC}{\affiliation{Astronomy, Astrophysics and Astrophotonics Research Centre, Macquarie University, Sydney, NSW 2109, Australia}}
\newcommand{\Capodimonte}{\affiliation{INAF - Capodimonte Astronomical Observatory, Salita Moiariello 16, I-80131 Napoli, Italy}}
\newcommand{\INFNNapoli}{\affiliation{INFN - Napoli, Strada Comunale Cinthia, I-80126 Napoli, Italy}}
\newcommand{\ICRANet}{\affiliation{ICRANet, Piazza della Repubblica 10, I-65122 Pescara, Italy}}
\newcommand{\MSU}{\affiliation{Center for Data Intensive and Time Domain Astronomy, Department of Physics and Astronomy,\\Michigan State University, East Lansing, MI 48824, USA}}
\newcommand{\IAP}{\affiliation{Institut d'Astrophysique de Paris, CNRS-Sorbonne Universit\'e, 98 bis boulevard Arago, 75014 Paris, France}}
\newcommand{\Pitt}{\affiliation{Department of Physics and Astronomy \& Pittsburgh Particle Physics, Astrophysics, and Cosmology Center (PITT PACC), University of Pittsburgh, 3941 O'Hara Street, Pittsburgh, PA 15260, USA}}
\newcommand{\Vtech}{\affiliation{Department of Physics, Virginia Tech, Blacksburg, VA 24061, USA}}
\newcommand{\IAC}{\affiliation{Instituto de Astrof{\'\i}sica de Canarias, E-38205 La Laguna, Tenerife, Spain}}
\newcommand{\Laguna}{\affiliation{Universidad de La Laguna, Dept. Astrof{\'\i}sica, E-38206 La Laguna, Tenerife, Spain}}
\newcommand{\UOak}{\affiliation{Homer L. Dodge Department of Physics and Astronomy, University of Oklahoma, 440 W. Brooks, Rm 100, Norman, OK 73019-2061, USA}}
\newcommand{\Hamburg}{\affiliation{Hamburger Sternwarte, Gojenbergsweg 112, D-21029 Hamburg, Germany}}
\newcommand{\PSI}{\affiliation{Planetary Science Institute, 1700 East Fort Lowell, Suite 106, Tucson, AZ 85719-2395 USA}}
\newcommand{\SETI}{\affiliation{SETI Institute, 339 Bernardo Ave, Suite 200, Mountain View, CA 94043, USA}}
\newcommand{\Hobart}{\affiliation{Physics Department, Hobart and William Smith Colleges, 300 Pulteney Street, Geneva, NY 14456, USA}}
\newcommand{\Cornell}{\affiliation{Department of Astronomy, Cornell University, 245 East Avenue, Ithaca, NY 14850, USA}}
\newcommand{\Athens}{\affiliation{IAASARS, National Observatory of Athens, Penteli 15236, Greece}}
\newcommand{\Turki}{\affiliation{Department of Physics and Astronomy, University of Turku, Vesilinnantie 5, 20500 Finland}}
\newcommand{\UMNAstro}{\affiliation{School of Physics and Astronomy, University of Minnesota, 116 Church Street S.E., Minneapolis, MN 55455, USA}}
\newcommand{\UVa}{\affiliation{Department of Astronomy, 530 McCormick Road, Charlottesville, VA 22904-4325, USA}}
\newcommand{\UTA}{\affiliation{Department of Physics, University of Texas at Arlington, Box 19059, Arlington, TX 76019, USA}}
\newcommand{\Konkoly}{\affiliation{Konkoly Observatory, CSFK, MTA Center of Excellence, Konkoly-Thege M. \'ut 15-17, Budapest, 1121, Hungary}}
\newcommand{\ELTE}{\affiliation{ELTE E\"otv\"os Lor\'and University, Institute of Physics and Astronomy, P\'azm\'any P\'eter s\'et\'any 1/A, Budapest, 1117 Hungary}}
\newcommand{\Szeged}{\affiliation{Department of Experimental Physics, University of Szeged, D\'om t\'er 9, Szeged, 6720, Hungary}}
\newcommand{\IAIFI}{\affiliation{The NSF AI Institute for Artificial Intelligence and Fundamental Interactions, USA}}
\newcommand{\UPadua}{\affiliation{Physics and Astronomy Department Galileo Galilei, University of Padova, Vicolo dell'Osservatorio 3, I-35122, Padova, Italy}}
\newcommand{\UWarwick}{\affiliation{Department of Physics, Gibbet Hill Road, University of Warwick, Coventry CV4 7AL, United Kingdom}}
\newcommand{\ING}{\affiliation{Isaac Newton Group of Telescopes, Apt. de Correos 368, E-38700 Santa Cruz de la Palma, Spain}}
\newcommand{\UNott}{\affiliation{School of Physics and Astronomy, University of Nottingham, University Park, Nottingham, NG7 2RD}}
\newcommand{\USurrey}{\affiliation{Mullard Space Science Laboratory, University College London, Holmbury St Mary, Dorking, Surrey RH5 6NT, United Kingdom}}
\newcommand{\USheffiled}{\affiliation{Department of Physics and Astronomy, University of Sheffield, Sheffield S3 7RH, UK}}
\newcommand{\Adler}{\affiliation{Adler Planetarium, 1300 S. DuSable Lake Shore Dr., Chicago, IL 60605, USA}}
\newcommand{\Monash}{\affiliation{School of Physics and Astronomy, Monash University, Clayton, Victoria 3800, Australia}}
\newcommand{\OzGrav}{\affiliation{OzGrav: The ARC Centre of Excellence for Gravitational Wave Discovery, Clayton, Victoria 3800, Australia}}
\newcommand{\Christ}{\affiliation{School of Physical and Chemical Sciences -- Te Kura Mat\={u}, University of Canterbury, Private Bag 4800, Christchurch 8140, \\ Aotearoa, New Zealand}}
\newcommand{\mta}{\affiliation{MTA-ELTE Lendület “Momentum” Milky Way Research Group, Szent Imre H. st. 112, 9700 Szombathely, Hungary}}
\newcommand{\bao}{\affiliation{Baja Astronomical Observatory of University of Szeged, Szegedi {\'u}t,
Kt. 766, 6500 Baja, Hungary}}
\newcommand{\hrs}{\affiliation{HUN-REN--SZTE Stellar Astrophysics Research Group, Szegedi {\'u}t, Kt.
766, 6500 Baja, Hungary}}

\author[orcid=0000-0003-4175-4960]{Conor~L.~Ransome}
\UA
\email[show]{cransome@arizona.edu} 

\author[0000-0003-4102-380X]{David J.\ Sand}
\UA
\email{dsand@arizona.edu}

\author[0000-0002-4924-444X]{K.\ Azalee Bostroem}
\Catalyst\UA
\email{bostroem@arizona.edu}

\author[0000-0002-7352-7845]{Aravind P.\ Ravi}
\UCD
\email{apazhayathravi@ucdavis.edu}

\author[0000-0001-8073-8731]{Bhagya M.\ Subrayan}
\UA
\email{bsubrayan@arizona.edu}

\author[0000-0003-0123-0062]{Jennifer E.\ Andrews}
\GeminiNorth
\email{jennifer.andrews@noirlab.edu}

\author[0009-0003-8380-4003]{Zachary G. Lane}
\Christ
\email{zachary.lane@pg.canterbury.ac.nz}

\author[0000-0002-7937-6371]{Yize Dong \begin{CJK}{UTF8}{gbsn}(董一泽)\end{CJK}}
\CfA
\email{yize.dong@cfa.harvard.edu}

\author[0000-0002-2028-9329]{Anya Nugent}
\CfA
\email{anya.nugent@cfa.harvard.edu}

\author[0000-0001-8818-0795]{Stefano Valenti}
\UCD
\email{stfn.valenti@gmail.com}

\author[0000-0002-0744-0047]{Jeniveve Pearson}
\UA
\email{jenivevepearson@arizona.edu}

\author[0000-0002-4022-1874]{Manisha Shrestha}
\Monash\OzGrav
\email{Manisha.Shrestha@monash.edu}

\author[0000-0002-1481-4676]{Samaporn~Tinyanont}
\affiliation{National Astronomical Research Institute of Thailand, 260 Moo 4, Donkaew, Maerim, Chiang Mai, 50180, Thailand}
\email{samaporn@narit.or.th}

\author[0000-0002-9454-1742]{Brian Hsu}
\UA
\email{bhsu@arizona.edu}


\author[0000-0002-1895-6639]{Moira Andrews}
\LCO\UCSB
\email{mandrews@lco.global}

\author[0009-0000-9929-7518]{Dominik B{\'a}nhidi} 
\Szeged\bao
\email{}

\author{Imre Barna B{\'i}r{\'o}} 
\bao\hrs
\email{}

\author[0000-0003-0528-202X]{Collin Christy}
\UA
\email{collinchristy@arizona.edu}

\author{Istv{\'a}n Cs{\'a}nyi}
\bao 
\email{}

\author[0000-0003-4914-5625]{Joseph Farah}
\LCO\UCSB
\email{jfarah@lco.global}

\author[0000-0003-4537-3575]{Noah Franz}
\UA
\email{nfranz@arizona.edu}



\author[0000-0003-2744-4755]{Emily T. Hoang}
\UCD
\email{emthoang@ucdavis.edu}

\author[0000-0002-0832-2974]{Griffin Hosseinzadeh}
\UCSD
\email{ghosseinzadeh@ucsd.edu}

\author[0000-0003-4253-656X]{D.\ Andrew Howell}
\LCO\UCSB
\email{ahowell@lco.global}

\author[0000-0003-0549-3281]{Daryl Janzen}
\USask
\email{daryl.janzen@usask.ca}


\author[0000-0001-8738-6011]{Saurabh W.\ Jha}
\Rutgers
\email{saurabh@physics.rutgers.edu}

\author[0000-0003-3108-1328]{Lindsey~A.~Kwok}
\CIERA
\email{lindsey.kwok@northwestern.edu}

\author[0000-0002-7866-4531]{Chang~Liu}
\NU
\CIERA
\SkAI
\email{ptg.cliu@u.northwestern.edu}

\author[0000-0001-9589-3793]{Michael Lundquist}
\Keck
\email{mlundquist@keck.hawaii.edu}

\author[0009-0001-3106-0917]{Aidan Martas}
\UCB \UCD
\email{aidmart@berkeley.edu}

\author[0000-0001-5807-7893]{Curtis McCully}
\LCO
\email{cmccully@lco.global}

\author[0009-0008-9693-4348]{Darshana Mehta}
\UCD
\email{ddmehta@ucdavis.edu}



\author[0000-0002-7015-3446]{Nicolas E.\ Meza Retamal}
\UCD
\email{nemezare@ucdavis.edu}

\author[0000-0001-5510-2424]{Nathan Smith}
\UA
\email{nathans@as.arizona.edu}

\author[0000-0003-4610-1117]{Tam\'as Szalai}
\Szeged\mta
\email{szaszi@titan.physx.u-szeged.hu}

\author[0000-0002-4951-8762]{Sergiy Vasylyev}
\UCSD
\email{svasylyev@ucsd.edu}

\author[0000-0002-5814-4061]{V.~Ashley~Villar}
\affiliation{Center for Astrophysics \textbar{} Harvard \& Smithsonian, 60 Garden Street, Cambridge, MA 02138-1516, USA}
\affiliation{The NSF AI Institute for Artificial Intelligence and Fundamental Interactions}
\email{ashleyvillar@cfa.harvard.edu}

\author[orcid=0009-0006-7296-728X]{Kathryn Wynn}
\LCO \UCSB 
\email{}




\begin{abstract}

Interacting supernovae probe the twilight years of massive stars, exhibiting signatures of interaction between the supernova ejecta and surrounding material expelled from the progenitor. We present the peculiar interacting supernova, SN\,2025ngs in NGC\,5961 (37.8\,Mpc). This transient toes the line between strongly interacting supernovae (type IIn) and type IIP supernovae. SN\,2025ngs presents photometrically as a short-plateau supernova, with a plateau duration, t$_{\mathrm{PT}}^{}\approx70$ days. Interaction features subside within a week post-explosion, consistent with the growing number of flash supernovae, giving way to a short period where a typical IIP spectrum is exhibited. Towards the drop off the plateau, interaction features re-emerge, exhibiting complex H$\alpha$ profiles throughout the rest of the transient evolution. We compare with models of early spectra, finding the abundances generally consistent with a supergiant progenitor with a high mass-loss rate (10$^{-3}$\,M$_\odot$\,yr$^{-1}$). Early, high-resolution spectra reveal a double-horned H$\alpha$ profile, providing strong evidence for shock interaction with a proximate disk-like circumstellar medium. Spectroscopically, SN\,2025ngs closely resembles the luminous SN\,1998S, despite photometric differences, with SN\,2025ngs having a relatively modest peak magnitude of $M_\mathrm{V}=-17.9$\,mag, adding another member to the surprisingly diverse 98S-like group.

\end{abstract}








\section{Introduction} 

When a massive star with an initial mass exceeding 8\,M$_\odot$ exhausts its nuclear fuel, it leads to a collapse terminating in an explosive death known as a core-collapse supernova \citep[CCSN][]{Woosley_2002}. CCSNe are classified primarily based on their spectral features \citep[e.g.][]{Filippenko_1997}, with hydrogen-rich CCSNe being type II SNe (SNe\,II), originating from progenitors that have retained some of their hydrogen envelope. SNe\,II are also split into photometric classes based on the morphology and timescale of the light curves, with SNe\,IIP exhibiting a recombination-driven plateau, and SNe\,IIL displaying a linear light curve decline, which is also recombination driven, but with a smaller hydrogen envelope \citep[albeit, the distinction blurs with large samples, indicating a continuum;][]{Anderson_2014}. Some SNe\,II exhibit spectral signatures that are the result of the ejecta interacting with dense, slow-moving circumstellar material (CSM) that was shed from the progenitor shortly before the terminal explosion \citep[see][for a review]{Smith_2017_review}. This mechanism manifests as narrow features, particularly on the hydrogen Balmer series. Strongly interacting SNe are classified as type IIn \citep[SNe\,IIn;][]{Filippenko_1989, Schlegel_1990, Filippenko_1997, Ransome_2021}. Some `normal' SNe exhibit narrow features (and often high-ionization lines) at early times which then fade relatively rapidly (within $\sim\,10$\,days). These features signify the interaction between the ejecta and a dense, confined CSM that is promptly swept up, and are sometimes known as `flash' SNe \citep[e.g.][]{Gal-Yam_2014,Khazov_2016, Kochanek_2019, Jacobson-Galan_2024}.

Interacting SNe, whether the `IIn-like' features are persistent, or fleeting, probe a critical, poorly understood period in the later life of massive stars. The observed progenitors of SNe\,IIP are red supergiants \citep[RSGs;][]{Smartt_2009_IIP}. However, the mass-loss rates inferred for SNe\,II with `flash' features are as high as $\leq10^{-1}$\,M$_\odot$\,yr$^{-1}$ in order to produce the $\sim0.1$\,M$_\odot$ of CSM that is observed \citep[e.g. SN\,2020tlf, SN\,2023ixf, and SN\,2024ggi;][]{Jacobson-Galan_2022, Jacobson-Galan_2023, Bostroem_2023, Shrestha_2024_24ggi}, while the canonical mass-loss rates of RSGs via supergiant winds are $\sim10^{-6}$\,M$_\odot$\,yr$^{-1}$ \citep[e.g.][]{Beasor_2020}. The mass-loss rates are even more extreme for the SNe\,IIn which range from 10$^{-3}-1\,$M$_\odot$\,yr$^{-1}$ \citep[e.g.][]{Hiramatsu_2024, Dickinson_2024, Dukiya_2024, Smith_2024, Ransome_2024}. While direct evidence is elusive, the mass-loss mechanisms may include the previously discussed massive winds, as well as unstable burning \citep[][]{Smith_2014}, gravity wave-driven pulsations \citep[][]{Shiode_2014, Wu_2022}, pair-instability pulsations \citep[][]{Woosley_2007, Woosley_2017, Woosley_2022}, and violent binary interactions \citep[][]{Kashi_2010,Kashi_2013,SmithArnett_2014,Schroder_2020, Ercolino_2024, Tsuna_2024}. Notably, in some cases, these mass loss events (which may be occurring in the months prior to the terminal explosion) are observable \citep[e.g. SN\,2009ip;][]{Berger_2009, Li_2009b, Smith_2010_09ip, Margutti_2014}. 




Moreover, there are transitional objects which lay in between SNe\,II and SN\,IIn. One example is SN\,1998S, which also had longer duration flash features than most other examples. This object photometrically evolved as a SN\,IIL, and was also one of the most luminous SNe known at the time of discovery \citep[peaking at $M_R^{}\approx-19.5$\,mag][]{Fassia_2000, Fassia_2001}. SN\,1998S is sometimes regarded as a SN\,IIn-L, referring to the linear light curve decline and the exhibited interaction features. However, SN\,1998S was not a `classic' SN\,IIn where the interaction manifested as an enduring, strong Lorentzian/Gaussian Balmer line profiles. Rather, after the initial interaction (i.e. flash) features subsided, the spectra evolved into a more standard SN\,II-like spectrum with broad, photospheric features, however, with complex line profiles. Curiously, the H$\alpha$ profile at around 70 days exhibited a multi-component profile shape, with no broad P-Cygni feature which would be typical of a standard SN\,II. The complex H$\alpha$ profile evolved through to the nebular phase at a few 100 days post-explosion. The spectral evolution of SN\,1998S revealed the presence of an extremely complex circumstellar environment formed by the mass loss of the progenitor \citep{Mauerhan_2012}. There are a number of other transitional objects such as PTF\,11iqb \citep[][]{Smith_2015}, SN\,2013fc \citep[][]{Kangas_2016}, and SN\,2024cld \citep{Killestein_2025}.


\begin{figure}[!ht]
    \centering
    \includegraphics[width=0.99\columnwidth]{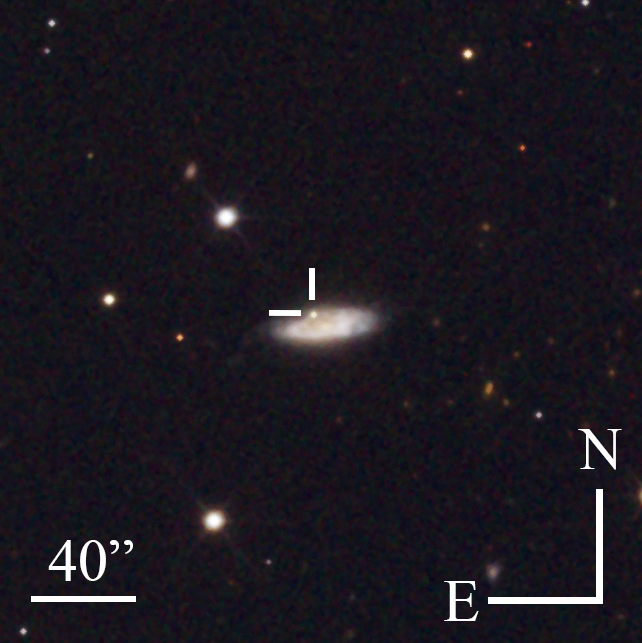}
    \caption{Color composite ($gri$ on 2025-07-31) Las Cumbres Observatory image of SN\,2025ngs and its host galaxy NGC\,5961.}
    \label{fig:host}
\end{figure}

 In this paper, we present ultraviolet-to-near infrared observations of the SN\,II, SN\,2025ngs which spectroscopically evolved similarly to SN\,1998S and related transients. SN\,2025ngs was discovered by the Asteroid Terrestrial-Impact Last Alert System \citep[ATLAS;][]{ATLAS} at 18.2\,mag in the ATLAS $o-$band \citep[][]{2025ngs_disc}, and was classified as a SN\,II by \citet{2025ngs_spec}. The host galaxy of SN\,2025ngs is NGC\,5961, at a distance of 37.8\,Mpc \citep[$\mu=32.89\,$mag;][]{Tully_2013} and is shown in Figure\,\ref{fig:host}. 



This paper is structured thusly: in Section\,\ref{sec:data} we outline our photometric and spectroscopic data obtained for SN\,2025ngs, and the associated data reduction procedures. We describe our analysis in Section\,\ref{sec:analysis}, including presenting the light curve, spectra, estimating the explosion time, probing the evolution of SN\,2025ngs, and comparing our data to models. We then discuss our results in relation to the possible progenitor scenarios for SN\,2025ngs in Section\,\ref{sec:prog}, characterize the host in Section\,\ref{sec:host}, and finally, we summarize our work in Section\,\ref{sec:conc}.

\section{Data} \label{sec:data}


\subsection{Photometry} \label{sec:phot}

Photometric observations of SN\,2025ngs were rapidly triggered post-discovery via the Global Supernova Project and Distance Less Than 40\,Mpc (DLT40) within a day of discovery. SN\,2025ngs was also promptly monitored with \textit{Swift} \citep[][]{Swift}. Our data and reduction procedures are summarized as follows:

\begin{enumerate}
    \item \textbf{DLT40:} The high cadence, targeted DLT40 survey searches for young transients in nearby galaxies using a network of the PROMPT telescopes \citep[see][for details on the reduction process]{Tartaglia_2017, Yang_2019}. Through DLT40, we obtained $BVgri$ and $Clear$ data. We also obtain $VR$ photometry from the Thai Robotic Telescope (TRT) network via the DLT40 project.
    \item \textbf{LCO:} We also have photometric data from the Las Cumbres Observatory network of telescopes \citep[][]{Brown_2013_lco}, via the Global Supernova Project in the $UBVgri-$bands. These data were reduced using the \texttt{LCOGTSNPIPE} pipeline based on \texttt{PyRAF} \citep[][]{Valenti_2016}, calibrated using the APASS and Landolt comparison catalogs (for $gri$ and $UBV$ filters, respectively). 
    \item \textbf{ATLAS:} Using the ATLAS forced photometry service \citep[][]{Tonry_2018, Smith_2020_atlas}, we obtain photometry in the ATLAS $c-$ and $o-$bands (cyan and orange), which are analogous to Pan-STARRS $g+r$ and $r+i$ filters, respectively. We then processed these forced photometry data using ATClean, for a final reduction of both the SN and pre-SN light curves \citep[][]{Rest_2025}.
    \item \textbf{Baja Astronomical Observatory:} SN\,2025ngs was also followed with the 0.8m BRC80 telescope found at the Baja Astronomical Observatory of the University of Szeged, Hungary. The instrumental {\it BVg'r'i'z'} magnitudes of the SN are calculated by the image subtraction method and are transformed into the standard photometric system \citep[for further technical details, see][]{Banhidi_2025}.
    \item \textbf{\textit{Swift:}} We obtained ultraviolet (UV) and optical data for SN\,2025ngs from the Ultraviolet/Optical Telescope \citep[UVOT,][]{Roming_2005} on the \textit{Neil Gehrels Swift Observatory} \citep[][]{Swift}. The first \textit{Swift} observations were on 13th June 2025. We monitored SN\,2025ngs with \textit{Swift} throughout the evolution of the transient in the optical $UBV$ filters, as well as in the UV, using the $UVW1$, $UVM2$, and $UVW2$ filters. These data were reduced using the High-Energy Astrophysics Software (HEASoft\footnote{\url{https://heasarc.gsfc.nasa.gov/docs/software/heasoft/}}) 
    with a circular source region (radius 3$\arcsec$) centered at the position of SN and background measured from a circular region (radius 5$\arcsec$) without contamination from other sources, we performed aperture photometry following standard UVOT analysis threads \footnote{\url{https://www.swift.ac.uk/analysis/uvot/}}. The zero-points for photometry were adapted from \cite{Breeveld11} with the latest time-dependent sensitivity corrections updated in 2020.
\end{enumerate}

\subsection{Spectroscopy} \label{sec:spec}

Rapid spectroscopic follow-up of SN\,2025ngs commenced almost immediately after discovery. The first optical spectrum was taken only a day post-explosion with GMOS on Gemini North  \citep[][]{Hook_2004}. Initially, we monitored SN\,2025ngs with high cadence spectroscopy using LCO, with a daily cadence for the early evolution of the transient. After this early phase, the cadence of our spectral observations were around 10 days. We also obtain spectra with Binospec on the MMT at early phases \citep[][]{Fabricant_2019_binospec}. Our spectroscopic observations are summarized in Table\,\ref{tab:spec}. We also have high resolution data from MAROON-X \citep[][]{maroonx} on Gemini-N. These high resolution echelle data are used in Section\,\ref{sec:ext} in order to estimate host extinction, using the host sodium absorption features, and in Section\,\ref{sec:csm} to analyze the H$\alpha$ profile. We also present NIR data from the NASA Infrared Telescope Facility (IRTF), and MMT MMIRS, with the MMIRS spectrum being our first spectrum for SN\,2025ngs.

Our spectra, and associated data-reduction procedures are summarized in Table\,\ref{tab:spec}, and are as follows:
\begin{enumerate}
    \item \textbf{Gemini:} Our first optical spectrum was obtained using GMOS on the 8.1\,m Gemini-N telescope \citep[][]{Hook_2004} on the 13th June 2025 (around a day post-explosion. In total, we have 11 GMOS spectra, using either the B480 or R400 grating, with the final GMOS spectrum being taken on the 13th Oct. 2025 (123 days post-explosion). These data were reduced using the Data Reduction for Astronomy from Gemini Observatory North and South pipeline \citep[\texttt{DRAGONS};][]{Labrie_2019}. As previously mentioned, we also have MAROON-X spectra taken on the 14th and 15th June (around two and three days post-explosion), reduced using the MAROON-X \texttt{DRAGONS}-based reduction pipeline \footnote{\url{https://github.com/GeminiDRSoftware/MAROONXDR}}.  
    \item \textbf{Las Cumbres Observatory:} Slightly after the first GMOS spectrum, the FLOYDS spectrograph \citep[on both the Las Cumbres Observatory Faulkes Telescope North and South 2\,m;][]{Brown_2013_lco} commenced our series of spectra via the Global Supernova Project. We have 12 FLOYDS spectra, with our last spectrum being taken on the 22nd Sept. 2025 (102 days post-explosion). These FLOYDS data were then reduced using the FLOYDS pipeline \citep{Valenti_2014}. 
    \item \textbf{MMT:} We have 2 spectra from Binospec \citep{Fabricant_2019_binospec} on the 6.5\,m MMT,  taken on  17th June 2025 and 18th June 2025 (5 and 6 days post-explosion). These Binospec data were reduced using the standard procedures in \texttt{PypeIt} \citep[][]{pypeit}. Our MMIRS data were reduced using standard procedures using an \texttt{IDL} based pipeline \citep[][]{Chillingarian_2015}.
    \item \textbf{IRTF:} We have four spectral epochs from the 3.2\,m NASA Infrared Telescope (IRTF). The first IRTF spectrum is from 24th June 2025 (around 12 days post-explosion), and the last IRTF spectrum was taken on 11th Sept. 2025 (95 days post-explosion). These spectra were reduced using the standard routines utilizing \texttt{SpeXtools} \citep[][]{Cushing_2004}, a similar procedure was followed for SN\,2024bch \citep[][]{Andrews_2025}.
    \item \textbf{MOSFIRE:} We have one spectral epoch from MOSFIRE \citep{MOSFIRE_2012} on the 10\,m Keck 1 telescope obtained on 7th Aug. 2025 (55 days post-explosion). The data contain three spectra covering J, H, and K bands. The data are reduced using the standard procedures in \texttt{PypeIt} \citep[][]{pypeit}.
\end{enumerate}

\section{Analysis} \label{sec:analysis}

In this section, we will outline our the analysis of our photometry and spectra.

\subsection{Extinction} \label{sec:ext}

The location of SN\,2025ngs within NGC\,5961 suggests that there may be significant line-of-sight extinction, as the transient is embedded in the disk of the relatively inclined host. As we have high resolution ($R\approx80,000$) spectra from MAROON-X, we can probe the Na\,ID absorption features in order to estimate the redshift of SN\,2025ngs, as well as the host line-of-sight extinction. The Na\,ID absorption due to the Milky Way is low signal-to-noise so in this work, we take the $E(B-V)_{\rm MW}^{}=0.025$ from \texttt{dustmaps} \citep[][]{Green_2018}, which uses the reddening maps of \citet[][]{Schlafly_2011}.

\begin{figure}
    \centering
    \includegraphics[width=0.48\textwidth]{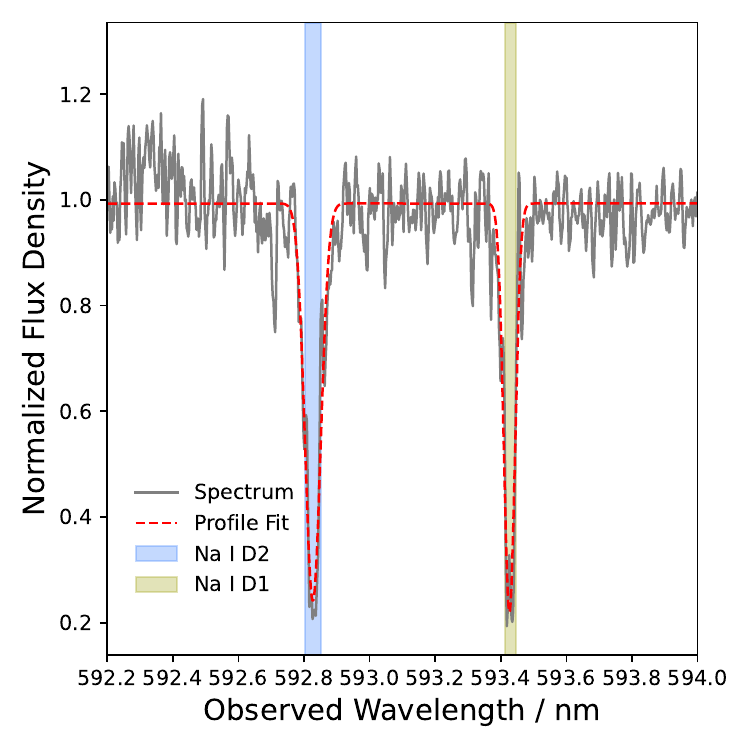}
    \caption{Our MAROON-X spectrum (we use our earliest spectrum from 14th Jun for this analysis) exhibiting the Na\,ID absorption due to the host. These absorption features are contained in the 103rd order of the blue side spectra. Both lines in the doublet are clearly resolved (data shown as a grey line). We show our double Gaussian fit to these profiles as a red dashed line. The regions representing the equivalent width of each feature are shaded.}
    \label{fig:na}
\end{figure}

Our high-resolution MAROON-X spectrum contains the Na\,ID absorption lines in the blue arm spectrum, in order 103. We use a Markov-Chain Monte Carlo (MCMC) fitting regime to fit Gaussian profiles to each of the sodium lines in the doublet. From these fits, we can infer the redshift of SN\,2025ngs, and also the host $E(B-V)$ using the empirical relation presented by \citet{Poznanski_2012} in their equation 9. These relations are dependent on the equivalent width of the absorption lines. We then infer a host $E(B-V)=0.152\pm0.018$\,mag \citep[with the uncertainty being from the fit, not including the dispersion from the empirical relations presented by][]{Poznanski_2012}, with $z=0.0065$. We show the region of our MAROON-X spectrum containing the Na\,ID absorption features in Figure\,\ref{fig:na}. We show the spectrum, our double Gaussian fit to the absorption profiles and also the regions representing the equivalent widths of each feature. This propagates to a host extinction of $A_V=0.471\pm0.056$\,mag, and a total extinction of $A_V = 0.549\pm0.056$\,mag.  This moderately high level of host extinction is consistent with SN~2025ngs's position within the host galaxy's disk.


\subsubsection{Photometric Evolution} \label{sec:ev}

\begin{figure*}
    \centering
    \includegraphics[width=0.95\textwidth]{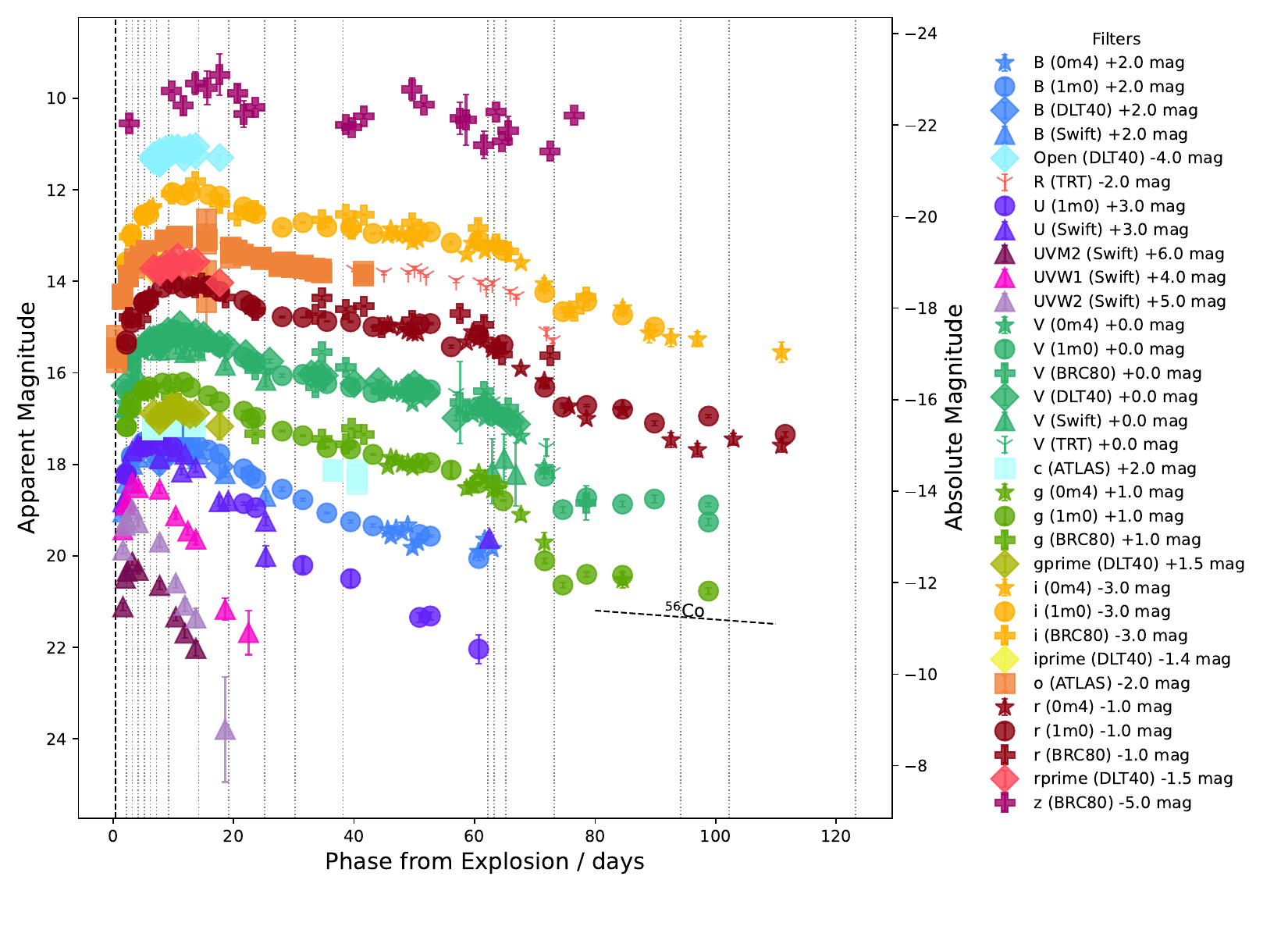}
    \caption{The light curve of SN\,2025ngs\footnote{Data behind the figure can be found on \href{https://zenodo.org/records/20056953}{Zenodo}}. We show our data spanning from MJD of 60838.4 to 60949.4 where SN\,2025ngs has faded below detection limits. Each telescope is assigned a different marker, our data include observations from LCO, DLT40, ATLAS, and \textit{Swift}. Our observations use the $UBVR$, $gri$, $g^{\prime}r^{\prime}i^{\prime}$, and \textit{Swift} $UVW1,W2,M2$ filters. These data are corrected for extinction, and are offset per filter. We mark our estimate of the explosion date with a black dashed line, and each spectral epoch by a gray dashed line. The evolution of SN\,2025ngs is relatively quick, with a short plateau before fading, followed by a slowly declining tail.}
    \label{fig:lc}
\end{figure*}

In Figure\,\ref{fig:lc}, we show the light curve of SN\,2025ngs over $\sim$120\,days of evolution post-explosion. We also mark the explosion time and the spectral epochs. The general light curve shape is typical of a SN\,IIP, albeit with a relatively short plateau phase \citep[e.g. SN\,2023ufx][with more examples within]{Ravi_2025}.

In order to estimate the explosion time, we take our ATLAS $o-$band data, which has constraining non-detections, and fit a parabolic rise to the early part of the light curve \citep[a common method to determine the explosion epoch. e.g.][see Figure\,\ref{fig:append}]{Nyholm_2020}. From this fit to the rise of SN\,2025ngs, we estimate the explosion time to be MJD\,$60837.8\pm0.2$. By interpolating our $V-$band light curve with a Gaussian process fit, we estimate the peak in the $V-$band to be at MJD $60848.5\pm0.1$, with a peak of $M_\mathrm{V}=-17.91\pm0.09$\,mag, giving us an $V-$band rise-time of $10.7\pm0.2$\,days. We summarize the basic properties of SN\,2025ngs in Table\,\ref{tab:properties}. SN\,2025ngs is luminous, being to 1$\sigma$ brighter than the mean in larger samples, such as the one presented by \citet{Anderson_2014}, which has a mean and spread of M$_\mathrm{V}=-16.74\pm1.01\,$mag. The rise time of $\sim11$\,days is similar to those in the sample presented by \citet[][]{Valenti_2016} and \citet{Gall_2015} who also show the rise times of SNe\,IIL, which are similar to SN\,2025ngs, but the brighter SNe have longer rises. The rise time of SN\,2025ngs is also similar to individual examples such as SN\,1998S \citep[][]{Liu_2000}, SN\,2023ixf \citep[][]{Hiramatsu_2023}, SN\,2023ufx \citep[][]{Ravi_2025, Tucker_2024}, and SN\,2024ggi \citep[][]{Shrestha_2024_24ggi}, but is rapid compared to the sample of strongly interacting SNe\,IIn presented by \citet{Ransome_2024}, where the median rise-time is around 40 days (caused by large amounts of CSM increasing the diffusion time).

\begin{figure}
    \centering
    \includegraphics[width=0.99\columnwidth]{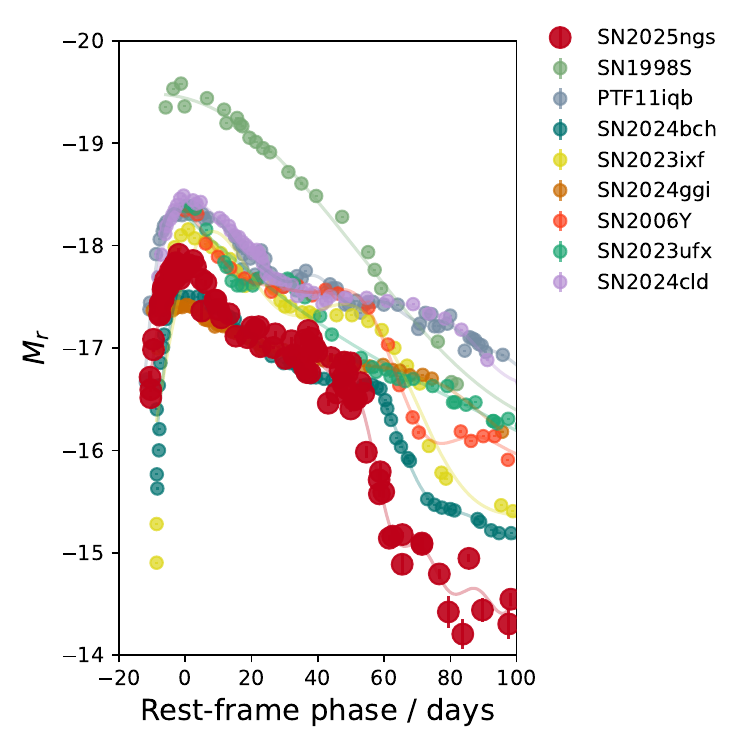}
    \caption{The $r/R/r^{\prime}-$band light curve of SN\,2025ngs and several comparison transients. These comparisons objects are SN\,1998S \citep[][]{Fassia_2000}, SN\,2006Y \citep{Hiramatsu_2021}, PTF11iqb \citep[][]{Smith_2015}, SN\,2023ixf \citep[][]{Hsu_2025}, SN\,2023ufx \citep{Ravi_2025}, SN\,2024bch \citep[][]{Andrews_2025}, SN\,2024cld \citep[][]{Killestein_2025}, and SN\,2024ggi \citep[][]{Ertini_2025}. Compared to most of the comparison objects, SN\,2025ngs has a short plateau, with a shape most similar to SN\,2024bch.}
    \label{fig:lccomp}
\end{figure}

After the rise to peak, SN\,2025ngs entered a plateau phase, exhibiting a SN\,IIP-like light curve morphology. The plateau levels out around 1\,mag fainter than the peak (e.g. in the $V-$band) which is a feature often attributed to the presence of CSM \citep[e.g.][]{Morozova_2018}. SNe\,IIP typically have a plateau duration of around 100\,days. We see that the plateau phase in the evolution of SN\,2025ngs is significantly shorter, indeed, the light curve has started to fall from the plateau around 60 days post-explosion. SN\,2025ngs therefore may be considered a member of the `short-plateau' SNe. \citep[e.g.][]{Tomasella_2013,Nakaoka_2019,Hiramatsu_2021, Tartaglia_2021,Teja_2022,Kilpatrick_2023_20jfo,Teja_2023_18gj,Teja_2024, Ravi_2025}. Similar to SN\,2025ngs, many of the members of the short plateau SN group are luminous when compared to the SN\,IIP population.

\begin{figure}
    \centering
    \includegraphics[width=0.99\columnwidth]{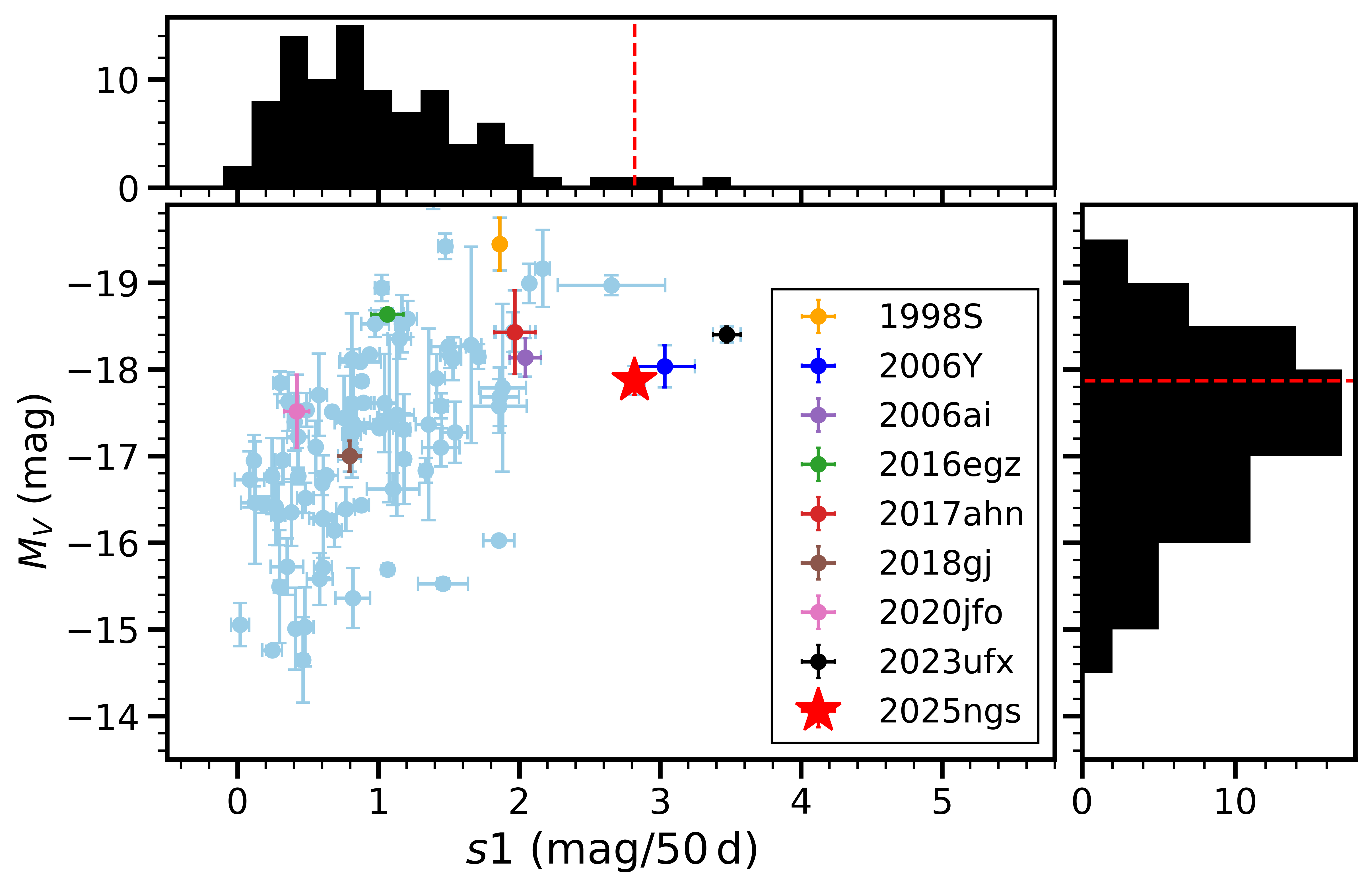}
    \includegraphics[width=0.99\columnwidth]{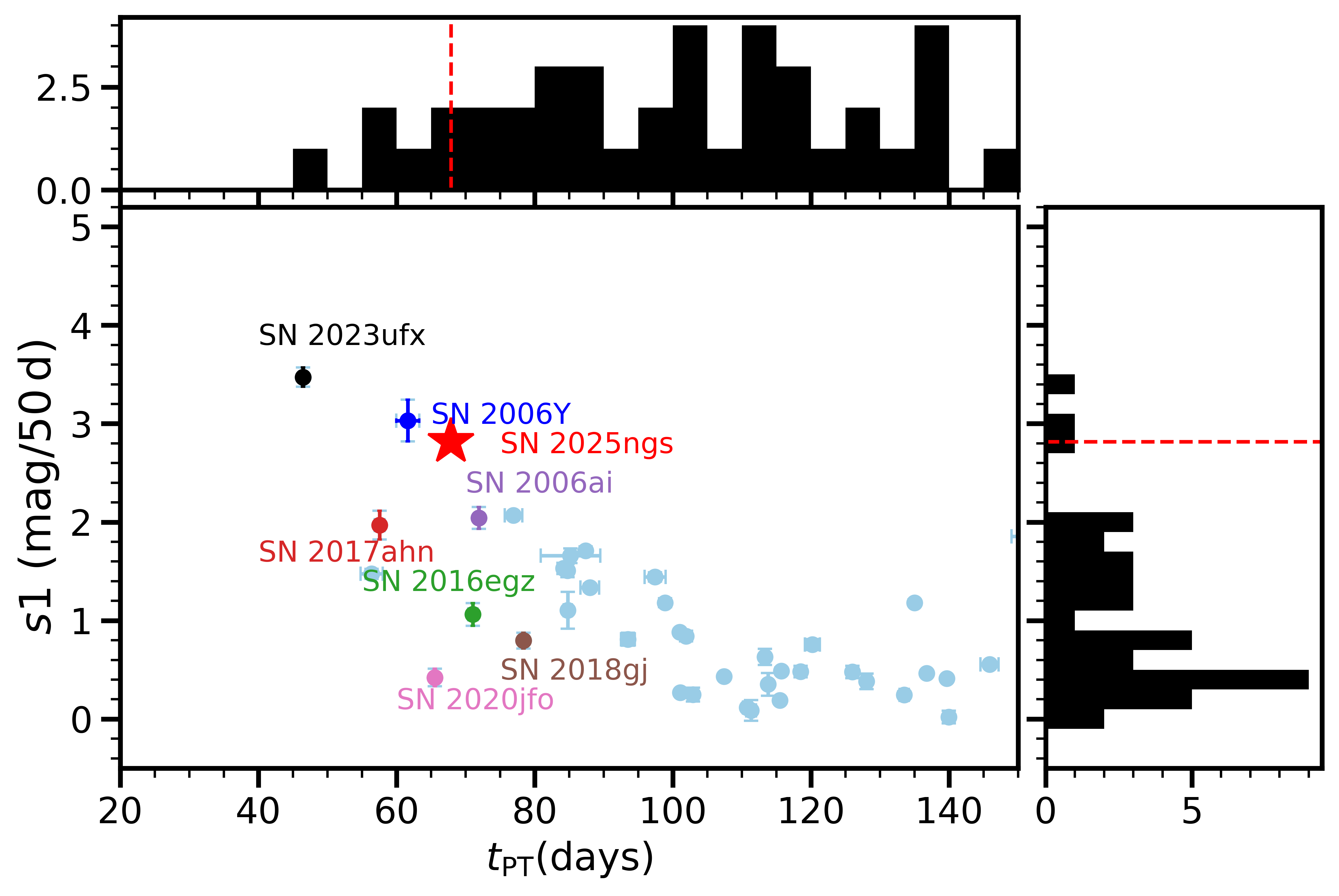}
    \caption{\textit{Top:} The distribution of the peak $V-$band absolute magnitude and the initial light curve decline, $s1$ for both a sample of SNe\,II from \citet{Anderson_2014}, and a selection of comparison objects, mostly short-plateau SNe\,II. The position of SN\,2025ngs is marked with a red star and we find that SN\,2025ngs is similar to the other short-plateau SNe\,2006Y and 2023ufx. \textit{Bottom:} The same as the previous figure but comparing $s1$ to the plateau duration.  SN\,2025ngs is again similar to SNe\,2006Y and 2023ufx, and is both faster declining than the main sample, and also the plateau duration is shorter than the majority of the comparison sample of SNe\,II.}
    \label{fig:mvs1}
\end{figure}

The $r/R-$band light curve of SN\,2025ngs compared to similar transients is presented in Figure\,\ref{fig:lccomp}. The plateau is shorter than the comparison objects, and is a similar peak brightness to the more weakly interacting SNe in this comparison set, such as SN\,2023ixf, and SN\,2024ggi. We compare the peak $M_V$ of SN\,2025ngs, a selection of short-plateau SNe\,IIP \citep[from][]{Ravi_2025}, SN\,1998S (as it is one of our comparison objects), and also the sample of SNe\,IIP presented by \citet{Anderson_2014}, \citet{Valenti_2016}, \citet{deJaeger_2019}, and \citet[][]{Anderson_2024} in Figure\,\ref{fig:mvs1}. We compare these peak luminosities to the rate of decay in the early decline, $s1$, measured in mag/50 days, which for SN\,2025ngs is 2.8\,mag/50d, calculated using the procedure presented by \citet{Valenti_2016}, and has been applied to previously presented samples \citep[e.g.,][]{Anderson_2014, Valenti_2016, deJaeger_2019, Anderson_2024}, as well as individual objects \citep[e.g. SN\,2017ahn and SN\,2023ufx][]{Tartaglia_2021,Ravi_2025}. From Figure\,\ref{fig:mvs1} we see that SN\,2025ngs has a steeper decline rate than the majority of the sample of SNe\,II. However, SN\,2025ngs has a similar $s1$ to other luminous short plateau SNe\,II such as SN\,2006Y \citep[see][]{Hiramatsu_2021}, and SN\,2023ufx. With an $M_V\approx-18$\,mag, SN\,2025ngs has a luminosity higher than the mean of the comparison sample (which has a mean peak $M_V\approx-16.7$\,mag). This is consistent with both the other luminous short plateau SNe\,II and also the correlation found by \citet{Anderson_2014} where the more luminous SNe\,IIP declined faster. The luminous SN\,1998S, however, had a somewhat slower decline than that of SN\,2025ngs, which may be due to a higher CSM mass compared to SN\,2025ngs. We then compare the $s1$ decline time to the plateau duration, $t_{\mathrm{PT}}^{}$ which is $\sim67$\,days. We present this parameter pair in Figure\,\ref{fig:mvs1}, again comparing with other short plateau SNe, and the larger SN\,IIP sample showing that SN\,2025ngs is similar in the plateau length to the other short plateau SNe, being on the short tail of the distribution which spans to around 150\ days.


At around 60 days post-explosion, the plateau ends, and a decline begins. This decline then levels out into a tail. In Figure\,\ref{fig:lc}, we display the gradient of the decay expected from radioactive decay. We fit a slope to the post-plateau tail (in the $V-$band) using an MCMC sampler (see Figure\,\ref{fig:append}). We find that the decline of the post-plateau tail is $0.0058\pm0.0024$\,mag\,day$^{-1}$. This slope is shallower than the decay expected for a tail powered by radioactive decay in the $V-$band ($0.0098$\,mag\,day$^{-1}$). This shallower gradient suggests that there is an additional power source to the light curve as opposed to radioactive decay being the sole engine. This is consistent with extra luminosity being provided by ongoing CSM interaction post-plateau. Indeed, the light curves of interacting SNe are often extended in duration due to CSM interaction  \citep[e.g., see][]{Ransome_2024}. Similarly, other SNe, such as SN\,2023ixf, lingering interaction with the extended wind has been observed to flatten out the late-time light curve \citep[e.g.][]{Jacobson-Galan_2025_23ixf}, however this is at a much later time than SN\,2025ngs, perhaps suggesting that there is significant underlying interaction in SN\,2025ngs. This behavior may suggest that after the plateau phase, as the SN photosphere recedes towards the core, the shocked CSM region is then revealed, and the luminosity from this coasting interacting region then flattens the light curve decline. Similar scenarios have been suggested for other interacting SNe such as SN\,1998S, PTF\,11iqb, and SN\,2024cld \citep[][]{Fassia_2001,Smith_2015,Killestein_2025}. As the post-plateau evolution has not yet reached a tail which is consistent with only radioactive decay, we can only place an upper limit on the nickel mass. Assuming complete trapping and the method from \citet{Hamuy_2003}, we get a limit of 0.007\,M$_\odot$.  We note that in terms of the photometric properties shown in Figure\,\ref{fig:mvs1}, SN\,2025ngs is most similar to SN\,2006Y and 2023ufx, with the SNe with proposed lower progenitor masses being closer to the main SN\,II population. It should be noted, however, that SN\,1998S is not placed close to SN\,2025ngs in Figure\,\ref{fig:mvs1}, but is more luminous than SN\,2025ngs and evolves as a SN\,IIL rather than a SN\,IIP. 



\begin{figure}
    \centering
    \includegraphics[width=0.99\columnwidth]{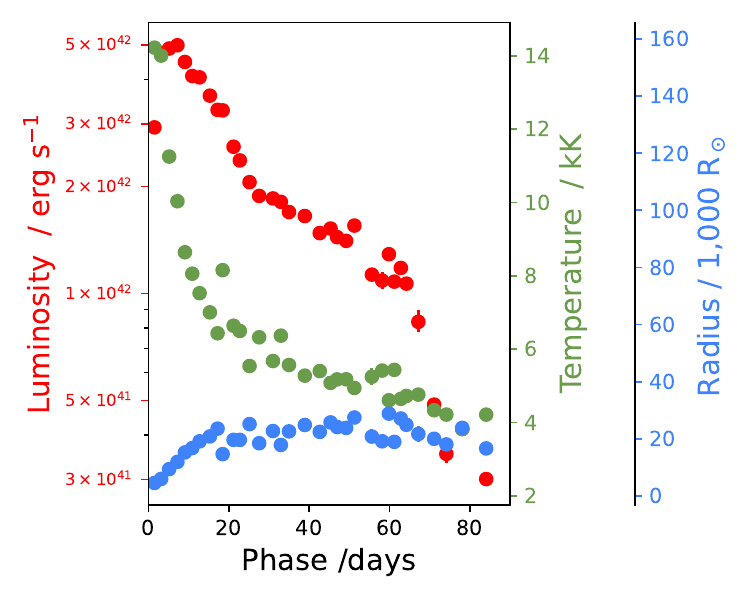}
    \caption{The black-body evolution of SN\,2025ngs, showing the luminosity, photospheric radius, and temperature.}
    \label{fig:bol}
\end{figure}

We use the Light Curve Fitting package \citep[][]{Hosseinzadeh_2023} to estimate the black-body evolution of SN\,2025ngs. We show the pseudo-bolometric light curve, along with the photospheric radius and temperature evolution in Figure\,\ref{fig:bol}. We find that SN\,2025ngs peaks at a few 10$^{42}$\,erg\,s$^{-1}$, lower than the $\sim10^{43}$\,erg\,s$^{-1}$ found for SN\,1998S \citep[]{Fassia_2000}, and the other short-plateau SNe \citep[e.g.][]{Hiramatsu_2021} which mostly have brighter peak $M_\mathrm{V}$. As is typical amongst SNe\,II, the temperature rapidly decreases post-shock breakout, leveling out at around 5,000\,K, the H recombination temperature \citep[see][]{Popov_1993, Valenti_2016, Faran_2018}. We also trace the radius of SN\,2025ngs, we see that there is a plateau at around the light curve plateau of $\sim2\times10^4$\,R$_\odot$, or around $10^{15}$\,cm. This has started to fall by the end of our observable window prior to the Sun-constraint. 

\begin{figure*}
    \centering
    \includegraphics[width=0.99\columnwidth]{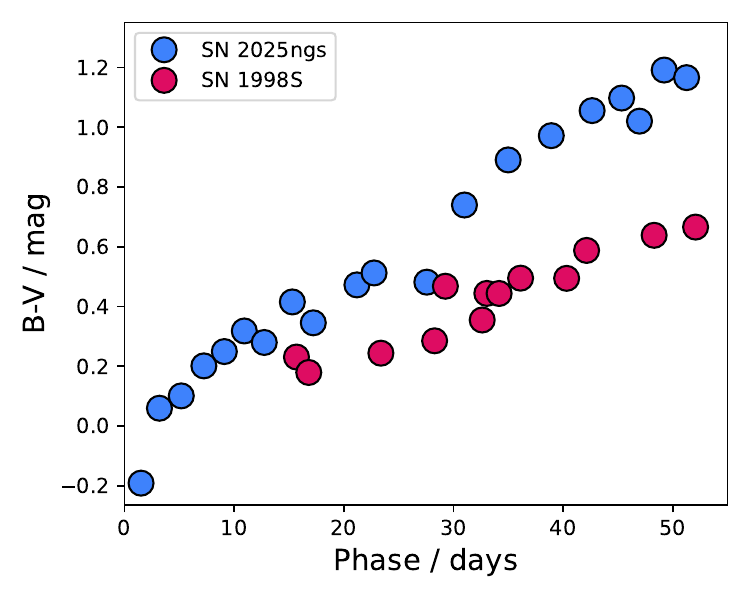}
    \includegraphics[width=0.99\columnwidth]{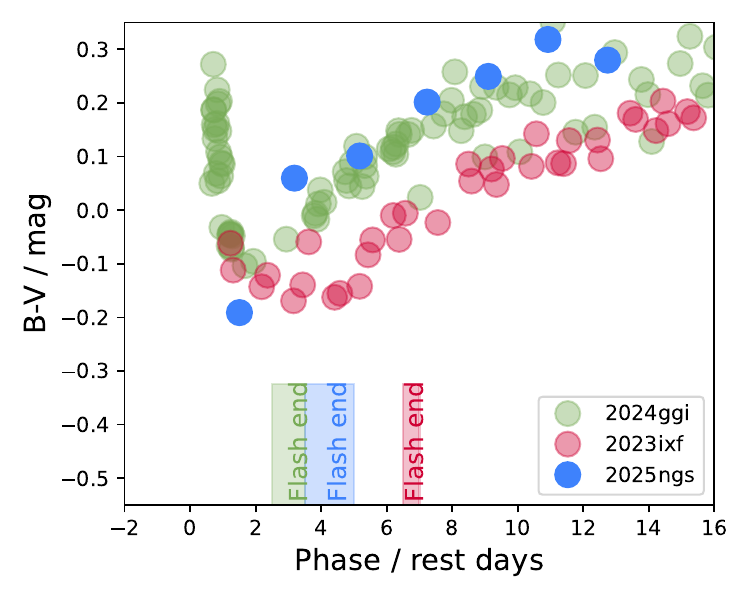}
    \caption{\textit{Left: }The $B-V$ color evolution of SN\,2025ngs compared to SN\,1998S \citep[data from][]{Fassia_2000}. \textit{Right:} The early $B-V$ color evolution of SN\,2025ngs, compared with the evolution of SN\,2023ixf and SN\,2024ggi \citep[from][]{Shrestha_2024_24ggi}. SN\,2025ngs exhibits a similar evolution to both SN\,2023ixf and SN\,2024ggi. }
    \label{fig:col}
\end{figure*}

Moreover, we compare the color-evolution of SN\,2025ngs with other, similar objects. Firstly, we present the color evolution of SN\,2025ngs compared against the spectroscopically similar, but photometrically distinct, SN\,1998S in Figure\,\ref{fig:col}. SN\,1998S does not have much early data, but we see that around 20 days post-explosion (around the time the plateau starts), both transients have a redward evolution in terms of their $B-V$ colors. As we have early photometry for SN\,2025ngs, we can also compare to the early color evolution of SN\,2023ixf and SN\,2024ggi, presented in Figure\,\ref{fig:col}. While we do not have the multi-band data at the earliest times to observe the very early phases showing the blueward evolution seen in SN\,2024ggi, and to a lesser extent, SN\,2023ixf, we do see a similar post-peak evolution in SN\,2025ngs. The overall color of SN\,2025ngs is most similar SN\,2024ggi, but has a bluer minima. This $B-V$ color may suggest underlying interaction, as is inferred in SN\,1998S \citep[][]{Leonard_2000}, albeit with the flash features in SN\,2025ngs not lasting as long as SN\,2023ixf \citep[][]{Bostroem_2023,Shrestha_2024_24ggi}. 

\begin{table}
    \centering
    \caption{Summary table of the basic properties of SN\,2025ngs.}
    \begin{tabular}{l|c}
        \hline \hline
         Parameter & Value\\
         \hline
         $\alpha$ (J2000)& 15:35:16.68   \\
         $\delta$ (J2000)& +30:51:55.73 \\
         $z$ & 0.0065 \footnote{from Na\,I\,D absorption.}\\
         Distance& 37.8\,Mpc \footnote{from \citet{Tully_2013} \label{fn:tully}}\\
         $\mu$& 32.89$\pm$0.20 \footref{fn:tully}\\
         Last Non-detection ($o-$band)& MJD 60837.4 \\
         First Detection (ATLAS $o-$band)& MJD 60838.4\\
         Explosion Time& MJD\,$60837.8\pm0.2$\\
         Peak $M_V$& $-17.91\pm0.09$\,mag\\
         $V-$band rise time& $10.9\pm0.2$\,days\\
         $E(B-V)_{\mathrm{MW}}^{}$& 0.025\,mag \\
         $E(B-V)_{\mathrm{host}}^{}$& $0.152\pm0.018$\,mag\\
         \hline \hline
    \end{tabular}
    \label{tab:properties}
\end{table}

\subsubsection{Shock-cooling Models} \label{sec:shock}

\begin{figure}
    \centering
    \includegraphics[width=0.95\columnwidth]{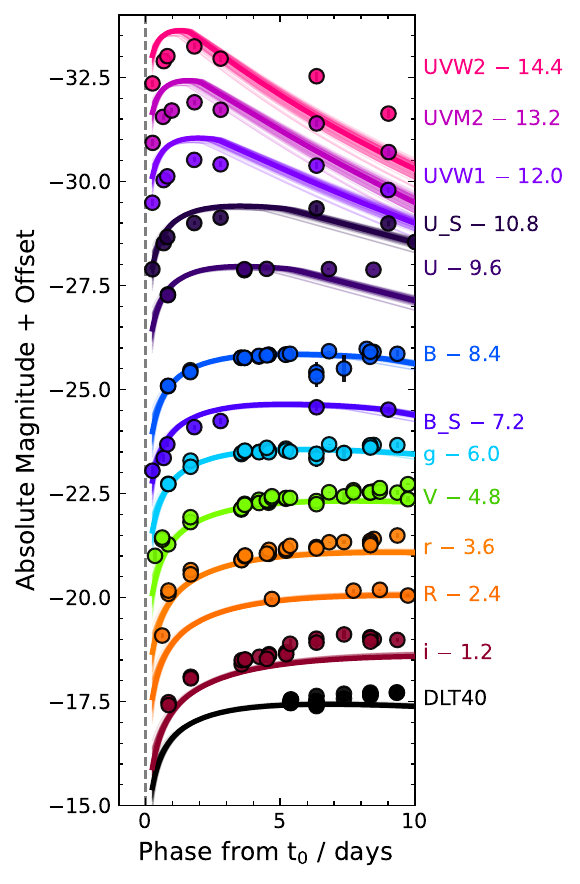}
    \caption{Our fits to the early light curve of SN\,2025ngs using the shock cooling models of  \citet{Morag_2023}, using Light Curve Fitting \citep[][]{Hosseinzadeh_2023}. We show the inferred explosion epoch as a grey dashed line, with each band and respective offsets labeled. The rise is not well fit in the bluer filters, indicative of a contribution to the early flux from CSM interaction.}
    \label{fig:cool}
\end{figure}

In core-collapse SNe, the early light curve may be shaped by shock-cooling emission. As we have early detections and high cadence photometric follow-up, with a well constrained rise time, we fit the early light curve of SN\,2025ngs to shock-cooling models \citep[e.g.,][]{Rabinak_2011,Sapir_2011, Katz_2012, Sapir_2013, Sapir_2017, Morag_2023, Morag_2024} using the Light Curve Fitting package \citep[][]{Hosseinzadeh_2020_lc, Hosseinzadeh_2023} and the model prescribed by \citet{Morag_2023}, which accounts for line-blanketing in the UV. These models have been applied to numerous SNe such as SN\,2021gmj \citep{Meza-Retamal_2024}, SN2022jox \citep[][]{Andrews_2024, Meza-Retamal_2024}, SN\,2023axu \citep[][]{Shrestha_2024_23axu}, SN\,2023ixf \citep[][]{Hosseinzadeh_2023}, SN\,2024ggi \citep[][]{Shrestha_2024_24ggi}, and samples \citep{Irani_2024}. 

We present our fit to the early light curve in Figure\,\ref{fig:cool}. We also record our prior ranges, sampling, and posterior median and 1$\sigma$ spread on the posterior distributions in Table\,\ref{tab:shock}. We use a similar set of prior distributions to those used for SN\,2023ixf \citep[][]{Hosseinzadeh_2023}, and SN\,2024ggi \citep[][]{Shrestha_2024_24ggi} for the shock velocity, envelope mass, the mass scaled by a numerical factor (of order unity), progenitor radius, explosion time, and intrinsic scatter to account for underestimated uncertainties in the photometry.

Comparing our inferred values to SN\,2024ggi \citep[][]{Shrestha_2024_24ggi}, the shock velocity is $\approx6800$\,km\,s$^{-1}$, consistent within the spread for the value inferred for SN\,2024ggi. The envelope mass is higher, and the progenitor radius is smaller. The explosion time is around 0.9 days later than our estimate from fitting a parabolic curve to the early light curve.

These shock cooling models do not fully capture the early light curve of SN\,2025ngs, particularly at bluer wavelengths, highlighting the need for early UV observations. This is similar to what was noted for other SNe that exhibited early CSM interaction, such as SN\,2023ixf \citep[][]{Hosseinzadeh_2023}, and SN\,2024ggi \citep[][]{Shrestha_2024_24ggi}. These poor fits are attributed to CSM interaction modifying the rise and shape of the early light curve, even though the rise time to peak is similar to other SNe\,IIP (also indicating that the inferred parameters be regarded with caution).

\begin{table*}
    \centering
    \begin{tabular}{l|ccc}
    \hline
         Parameter&Prior Range  & Sampling Function & Median and Spread \\
         \hline \hline
         Shock Velocity, $v_{s*}$ ($\times10^{8.5}$\,cm\,s$^{-1}$) & $1-5$  &Uniform  &2.14$^{+0.09}_{-0.07}$ \\
         Envelope Mass, $M_{\mathrm{env}}$ ($M_\odot$) & $0-10$  & Uniform  &3.30$^{+2.6}_{-0.7}$ \\
         Scaled Ejecta Mass, $f_pM$ ($M_\odot$) & $0.05-100$ & Uniform  &0.82$^{+0.14}_{-0.08}$ \\
         Progenitor Radius,$R$ ($\times10^{13}$\,cm)& $0-2000$ & Uniform  &1.70$^{+0.20}_{-0.20}$ \\
         Explosion Time, $t_0$ (MJD)& $60836-60842$  & Uniform  &60838.69$^{+0.07}_{-0.18}$ \\
         Intrinsic Scatter, $\sigma$& $0-100$ & Log-uniform & 4$^{+2}_{-2}$ \\
         \hline \hline
    \end{tabular}
    \caption{Shock-cooling model parameters, prior ranges, sampling, and posterior distribution mean values presented with the 16th and 84th percentile spreads for each parameter.}
    \label{tab:shock}
\end{table*}

\subsubsection{Searching for Precursor Emission} \label{sec:pre}

The presence of CSM  implies enhanced mass loss shortly prior to the terminal explosion. These mass loss episodes may be accompanied by observable outbursts, which are common in SNe\,IIn \citep[e.g.][]{Ofek_2014, Strotjohann_2020}. Such emission has only been seen in a single normal SN\,II, SN\,2020tlf \citep[][]{Jacobson-Galan_2022}, despite extensive searches for precursor activity in very nearby SNe with exquisite observations and long-baseline pre-explosion datasets, for example SN\,2023ixf \citep[][]{Dong_2023, Ransome_2024_23ixf}, and SN\,2024ggi \citep[][]{Shrestha_2024_24ggi}. 

We explore the ATLAS pre-explosion dataset, which spans over 2500 days before the explosion time of SN\,2025ngs in both the ATLAS $o-$ and $c-$bands. We process the ATLAS data using \texttt{ATClean} \citep[][]{Rest_2025}, filtering out flagged data, and using 5$\sigma$ upper limits. We do not find evidence of precursor emission from the progenitor of SN\,2025ngs down to around $-12$\,mag in both filters. This is within the luminosity region inhabited by LBV outbursts. There are isolated, spurious detections in the pre-explosion dataset (in the $o-$band), but these are concurrent with non-detections.




\subsection{Spectroscopic Evolution} \label{sec:specev}

\begin{figure*}
    \centering
    \includegraphics[width=0.95\textwidth]{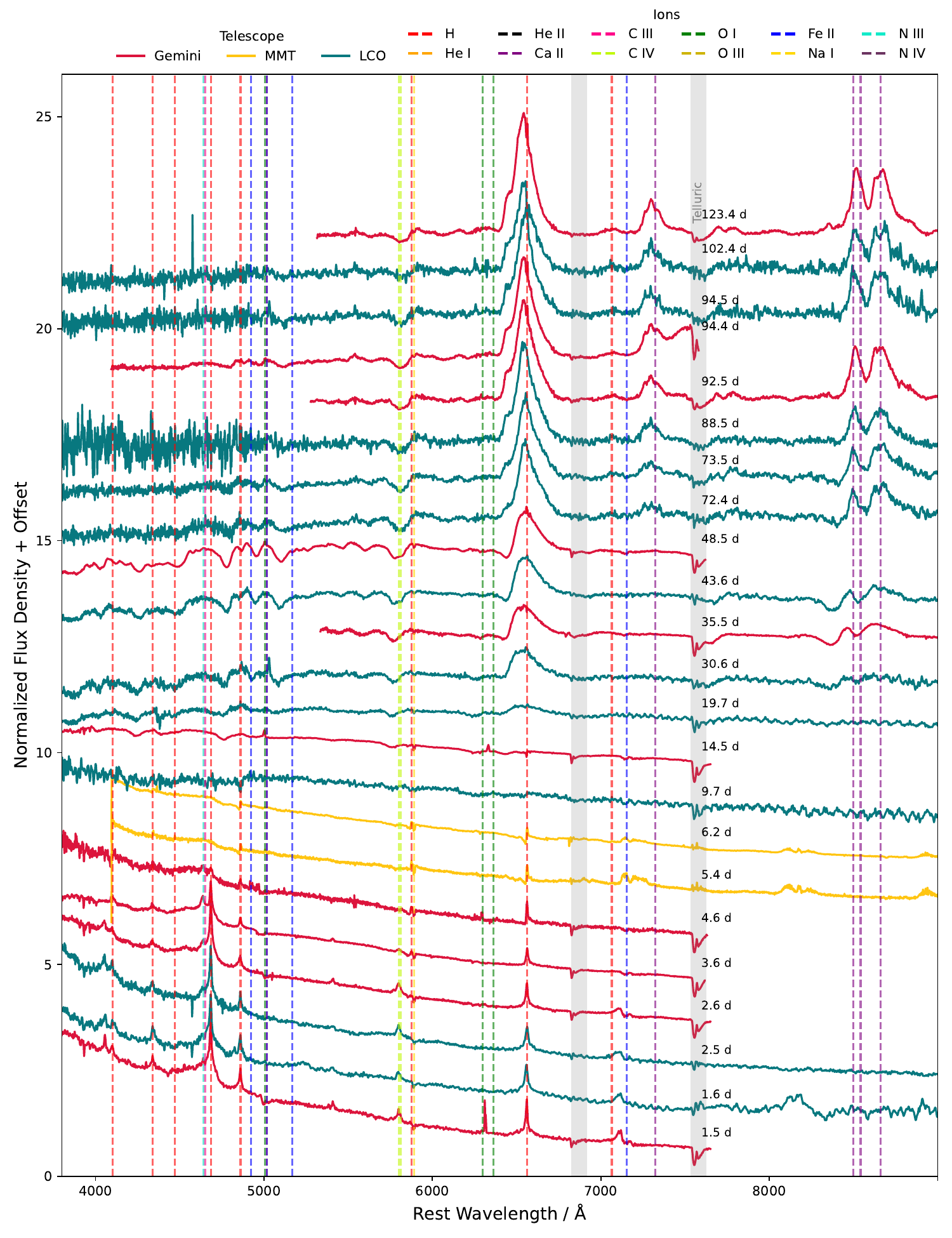}
    \caption{The spectral time-series of our optical data of SN\,2025ngs\footnote{Data behind the figure can be found on \href{https://zenodo.org/records/20056953}{Zenodo}}. The sequence starts at the bottom of the panel, with the phase relative to the explosion time labeled. We mark transitions of interest, such as lines associated with early interaction (flash), and the Balmer series with vertical dashed lines. Regions where telluric lines are present are shaded in grey. }
    \label{fig:spec}
\end{figure*}

Our spectra range from 13th June 2025 (around a day post-explosion) to the 13th October 2025 (around 123 days post-explosion). We show our full spectral sequence in Figure\,\ref{fig:spec}. We also illustrate the evolution of the H$\alpha$ profile shifted into velocity space on the right panel of Figure\,\ref{fig:spec}. We then show our time-series compared to comparable epochs of similar SNe in Figure\,\ref{fig:spec_comps}. 

\subsubsection{Early Evolution}

Our first optical spectrum, obtained with GMOS on Gemini-N around a day post-explosion exhibits a blue continuum along with Balmer series emission and high ionization lines such as C\,III, N\,III, N\,IV, and C\,IV. These features are characteristic of early time interaction with very proximate CSM which must have been ejected via some mechanism by the progenitor shortly before the terminal explosion  \citep[i.e. the previously discussed flash features][]{Khazov_2016, Kochanek_2019, Jacobson-Galan_2024}. These flash features are apparent in the first 5 spectra, extending to the 15th June, and have mostly disappeared by around 6 days post-explosion. Our seventh spectrum from Binospec on MMT transitions to a more featureless regime, with Balmer lines (H$\alpha$ and H$\beta$) being the only strong features \citep[but the complex profiles still indicate  CSM interaction, similar to what was seen in SN\,2024bch;][]{Andrews_2025}. This flash phase indicates when the SN ejecta has overtaken this nearby CSM, engulfing the shock interaction. Taking the ejecta velocity to be around 7,000\,km\,s$^{-1}$ (at around 40 days post-explosion), estimated from the offset of the trough minima of the P-Cygni profile seen in later epochs, we can place a lower limit on the extent of the confined CSM to $\approx3\times10^{14}$\,cm, largely consistent with other SNe that exhibit early narrow emission features, such as SN\,2023ixf \citep[][]{Jacobson-Galan_2023, Bostroem_2023}, SN\,2024bch \citep[$2.4-3.6\times10^{14}$\,cm;][]{Andrews_2025}, and SN\,2024ggi \citep[$\sim3.1\times10^{14}$\,cm][]{Shrestha_2024_24ggi}. 

\begin{figure*}[!ht]
    \centering
    \includegraphics[width=0.95\textwidth]{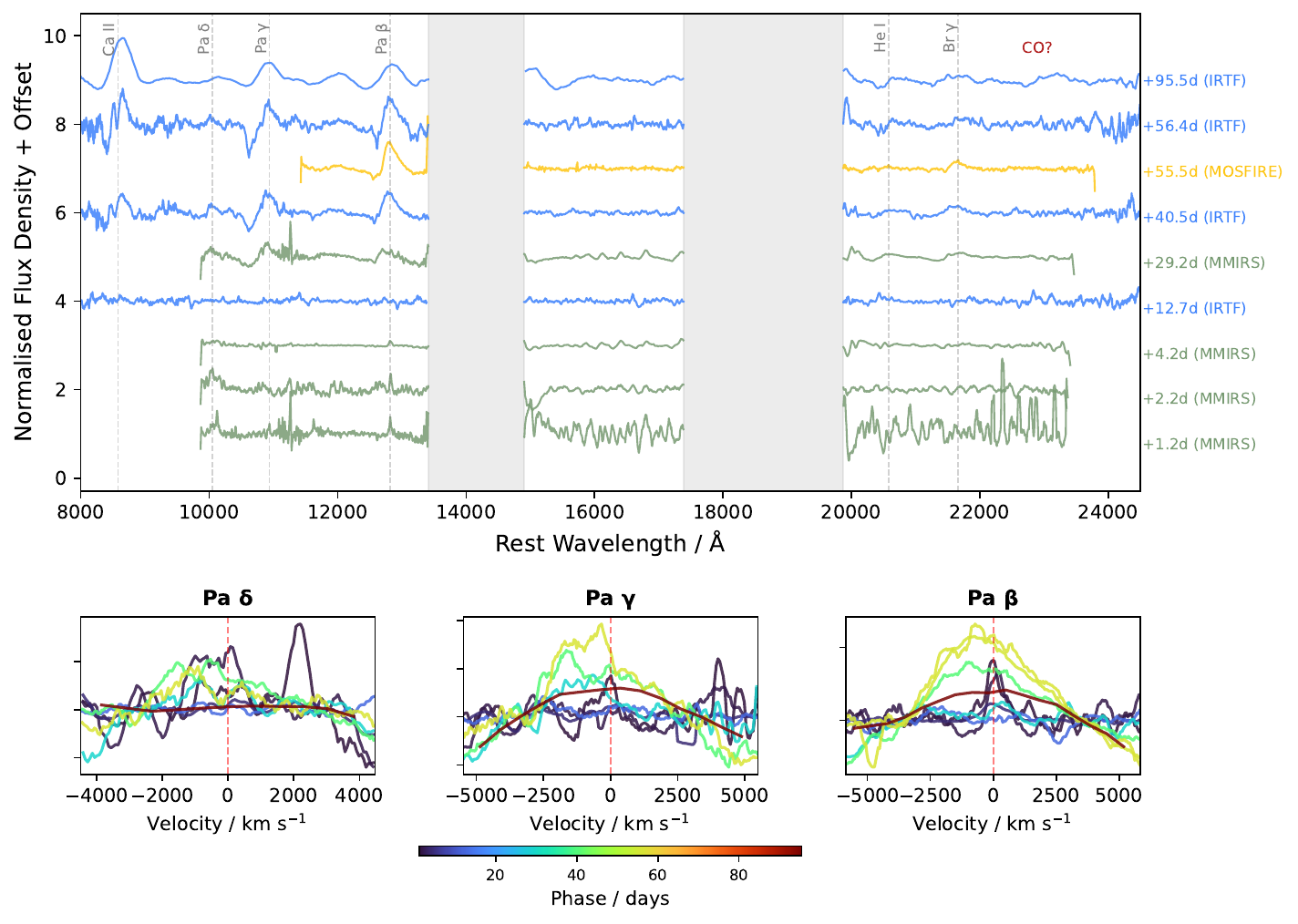}
    \caption{The near-infrared spectral time-series of SN\,2025ngs. We show the smoothed spectra alongside unsmoothed spectra. Telluric regions are shaded gray. These spectra span from 13th June (around a day post-explosion) to 11th July.  For readability, each spectrum is scaled and offset. The bottom panels show zoom-ins of the Paschen series lines, showing the fading of the flash lines. }
    \label{fig:mmirs}
\end{figure*}

We present a time series of our NIR data in Figure\,\ref{fig:mmirs}.  This includes early NIR data with MMIRS, taken around a day post-discovery. The early NIR data\footnote{Triggered using the PyMMT API wrapper \citep[][]{pymmt,Shrestha_2024_23axu}} reveal flash features, such as Pa\,$\beta$, Pa\,$\gamma$ and He\,II.  These spectra cover two early epochs (one day, and four days post-discovery), and also a later spectrum from around a month post explosion. We see the flash features, along with a blue continuum clearly on the first spectrum, covering the $zJ, H, K-$bands. The next spectrum from 3 days later shows the blue continuum shape persisting, but most of the flash features have subsided, coincident with what is seen in the optical spectral time-series. 


\begin{figure}
    \centering
    \includegraphics[width=0.98\columnwidth]{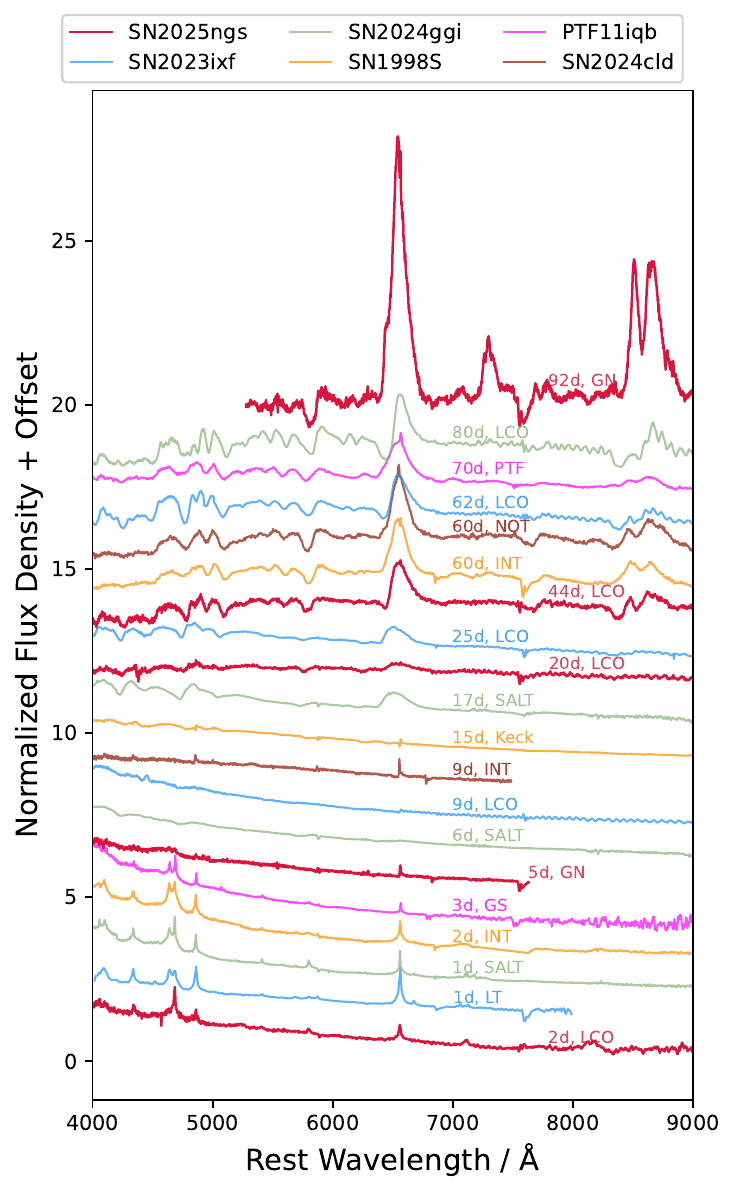}
    \caption{Selected phases from our time-series of SN\,2025ngs compared to SNe that are similar, at least at some epoch. We select SN\,1998S \citep[][]{Fassia_2001}, PTF\,11iqb \citep[][]{Smith_2015}, SN\,2023ixf \citep[][]{Jacobson-Galan_2023, Hosseinzadeh_2023}, SN\,2024cld \citep[][]{Killestein_2025}, and SN\,2024ggi \citep[][]{Shrestha_2024_24ggi}, as our comparison spectra. We also note the phase and the telescope used.}
    \label{fig:spec_comps}
\end{figure}

\begin{figure*}[!ht]
    \centering
    \includegraphics[width=0.95\textwidth]{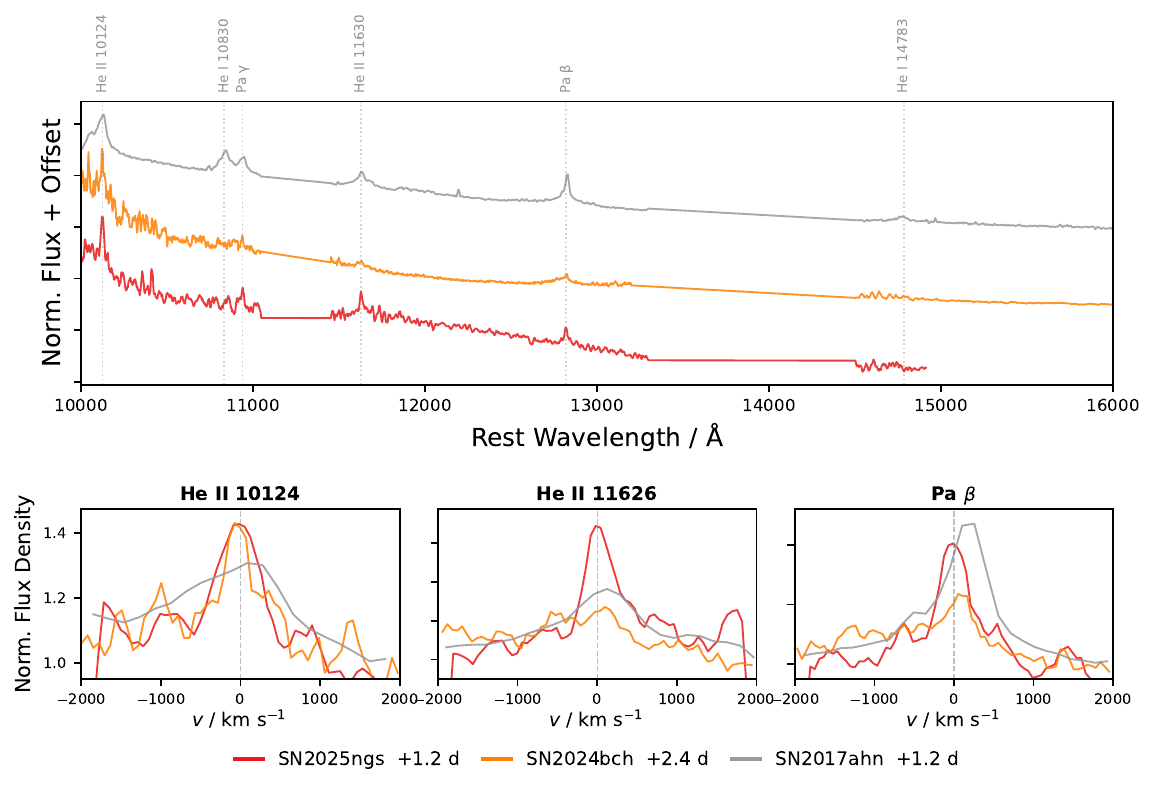}
    \caption{A comparison of the early NIR spectra of SN\,2025ngs (MMIRS), SN\,2024bch \citep[MMIRS, from][]{Andrews_2025}, and SN\,2017ahn \citep[Flamingos2, from][]{Tartaglia_2021}.  There are clear narrow He\,II lines, as well as Pa\,$\beta$. We also present zoom-ins on the clear He\,II lines and Pa\,$\beta$ in the bottom panels. The earliest spectrum for SN\,2024bch shows weaker flash lines, perhaps indicating that these features fade quickly at NIR wavelengths.}
    \label{fig:nirflash}
\end{figure*}

The detection of NIR flash features is rare. This may be due to a paucity of early NIR spectra, or perhaps these features fade quickly at longer wavelengths. This may be seen in Figure\,\ref{fig:nirflash}, where we present a comparison with the other observed NIR flash features, from SN\,2017ahn \citep[][]{Tartaglia_2021_17ahn}, and SN\,2024bch \citep[][]{Andrews_2025}. The flash features we see in SN\,2025ngs are also clear in SN\,2017ahn at a similar epoch, with Paschen series and helium features being apparent. Such features are also seen in the spectrum of SN\,2024bch, however these are muted, with only the strongest flash features being clearly detected (Pa\,$\beta$ and He\,II $\lambda$10124). The MMIRS spectrum of SN\,2024bch was taken at an epoch over a day after that of SN\,2017ahn and SN\,2025ngs. While the optical flash features persisted for around 6 days post-discovery, the NIR features have mostly faded in the case of SN\,2024bch by 2.4\,days \citep[][]{Andrews_2025}, so the dearth of observed flash features in the NIR may also be due to these features fading quickly. We focus on some of the key NIR flash features in the bottom panels of Figure\,\ref{fig:nirflash}. We see that these lines exhibit the classic Lorentzian profile shapes centered at 0\,km\,s$^{-1}$, as is expected from these flash features (due to electron scattering). We compare these early features to similar SNe in Figure\,\ref{fig:spec_comps}. There are differences in the velocities seen in the flash phases of SN\,2025ngs and the comparison objects. This may be an effect of the slightly different phases between each spectrum and the rapidly evolving nature of these features, an intrinsic difference between each SN (for example, higher ejecta velocities), or differences in resolution of each instrument. Over the first week or so, all of these comparison objects exhibit high ionization flash features. 


\subsubsection{High-resolution Spectroscopy}

\begin{figure*}
    \centering
    \includegraphics[width=0.98\columnwidth]{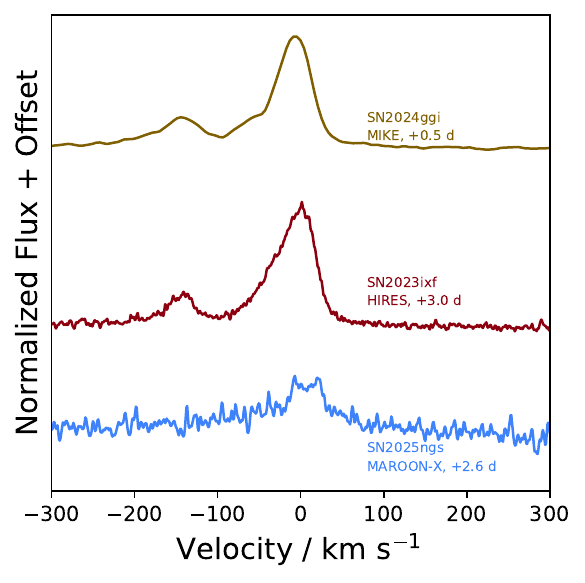}
    \includegraphics[width=0.98\columnwidth]{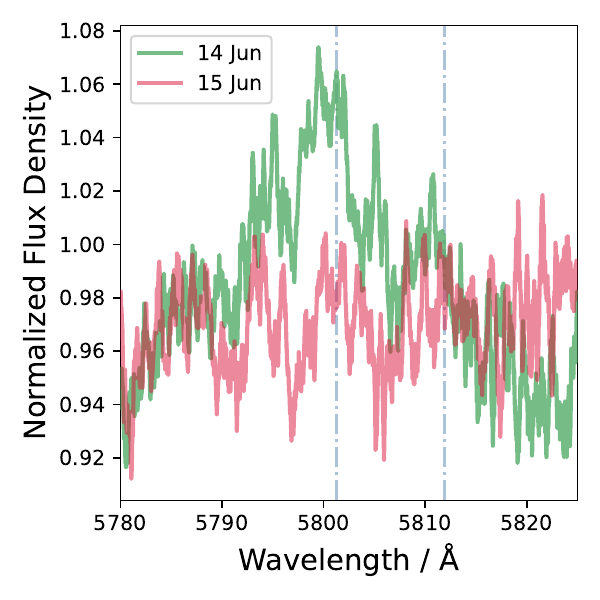}
    \caption{\textit{Left:} The early H$\alpha$ regions from high-resolution spectra compared between SNe\,2023ixf, 2024ggi, and 2025ngs. The SN\,2023ixf spectrum was taken from \citet{Dickinson_2025}, and the SN\,2024ggi spectrum is from \citet[]{Shrestha_2024_24ggi}. \textit{Right:} Zoom-in on the C\,IV lines in our early MAROON-X spectra. These features fade after one day, as also seen by \citet{Shrestha_2024_24ggi} in SN\,2024ggi. We mark the positions of the C\,IV (at 5801\AA\, and 5811\AA\, respectively) features as dashed lines. }
    \label{fig:earlycomp}
\end{figure*}

\begin{figure}[h]
    \centering
    \includegraphics[width=0.99\columnwidth]{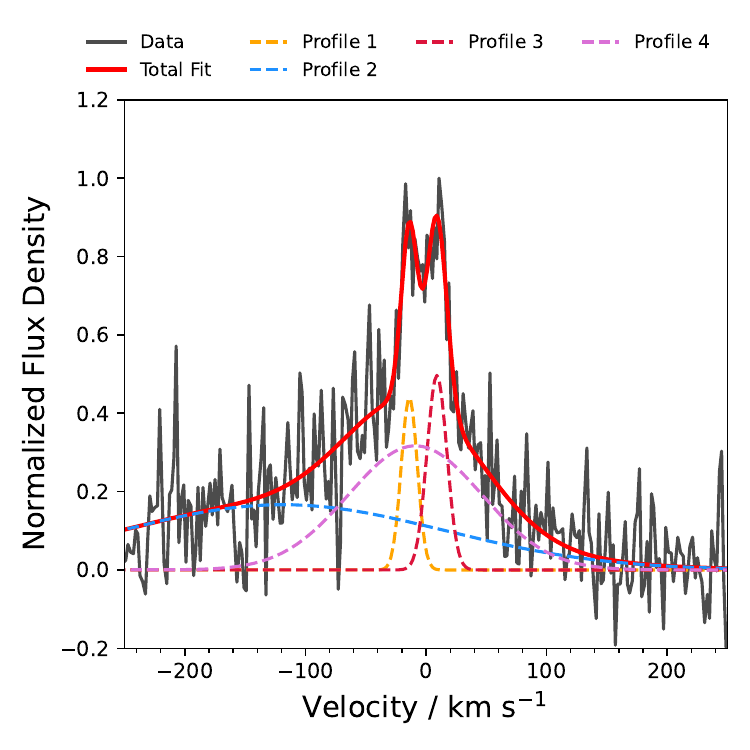}
    \caption{The H$\alpha$ of SN\,2025ngs as observed by MAROON-X on Gemini North. This profile has been shifted into velocity space, with normalized flux density. We show the original profile, without continuum subtraction, and also the subtracted profile with Gaussian fits applied to the complex profile. This spectrum was taken around 2 days post-explosion, and exhibits a complex shape, with electron scattering wings (fit with the blue dashed line), as well as a double peaked structure, indicative of interaction with a circumstellar disk.}
    \label{fig:maroonx}
\end{figure}

High resolution spectra from MAROON-X on Gemini North \citep[][]{Seifahrt_2016, Seifahrt_2018} were taken on 14th and 15th June 2025 (corresponding to around two and three days post-explosion). MAROON-X has R\,$\geq80,000$, or a resolution of order a few km\,s$^{-1}$. These data are therefore able to probe the winds of RSGs, with canonical wind velocities of a few 10s km\,s$^{-1}$. We used these data in Section.\,\ref{sec:ext} to probe the interstellar sodium absorption from the host, NGC\,5961, in order to estimate the redshift. In Figure\,\ref{fig:earlycomp}, we show the early spectra of SN\,2025ngs, SN\,2024ggi, and SN\,2023ixf. These objects have early high resolution spectra, we show the concurrent lower-resolution spectrum on the left panel, along with the corresponding H$\alpha$ profile from the high-resolution spectrographs. The high resolution H$\alpha$ profiles of SN\,2024ggi \citep[from HRS on SALT;][]{Shrestha_2024_24ggi}, and SN\,2023ixf \citep[from NEID on WIYN;][]{Dickinson_2025}, and SN\,2025ngs are presented in the right panel of Figure\,\ref{fig:earlycomp}. The H$\alpha$ profiles of SN\,2024ggi and SN\,2023ixf both show blue wing, and a He\,II feature at around --150\,km\,s$^{-1}$. This is in contrast with SN\,2025ngs, which does not show the helium feature (however, our MAROON-X spectrum is relatively low S/N, so we can not rule out that this helium feature is present). Rather than the classic sharp peak usually seen in the H$\alpha$ profile, we see a flatter topped profile with two peaks. This feature is shown in Figure\,\ref{fig:maroonx}, and will be discussed further in Section\,\ref{sec:csm}, and may suggest that the progenitor system of SN\,2025ngs may be distinct from the weaker interacting SNe\,II. The high resolution data for SN\,2024ggi revealed narrow, rapidly evolving C\,IV lines ($\lambda5801$, and $\lambda5811$, which form an apparent double profile), rarely seen in flash SNe \citep[notably seen in SN\,2023ixf and SN\,2024ggi][]{Smith_2023_ixf,Shrestha_2024_24ggi}. We present the MAROON-X data at these wavelengths in Figure\,\ref{fig:earlycomp}. We see that, while the signal-to-noise ratio is not as high as the data obtained by \citet{Shrestha_2024_24ggi}, as SN\,2024ggi was only at $\approx7$\,Mpc, we do see these carbon features in the earliest spectrum (taken on 14th June 2025). These features have subsided by the spectrum taken on 15th June 2025. \citet[][]{Shrestha_2024_24ggi} note that in their spectrum of SN\,2024ggi, these high ionization carbon features increase in strength between their epochs (separated by around 4 hours), indicating an increase in temperature (and ionization fraction), this behavior was also seen in SN\,2023ixf \citep[][]{Smith_2023_ixf}. In contrast, our data shows a fading of these features, indicating cooling. We do note, however, that our high resolution spectra are at a later relative phase, so may not capture the initial heating of the confined CSM, rather, only the cooling and subsequent fading of this feature.

\subsubsection{Later Evolution}

\begin{figure*}
    \centering
    \includegraphics[width=0.8\textwidth]{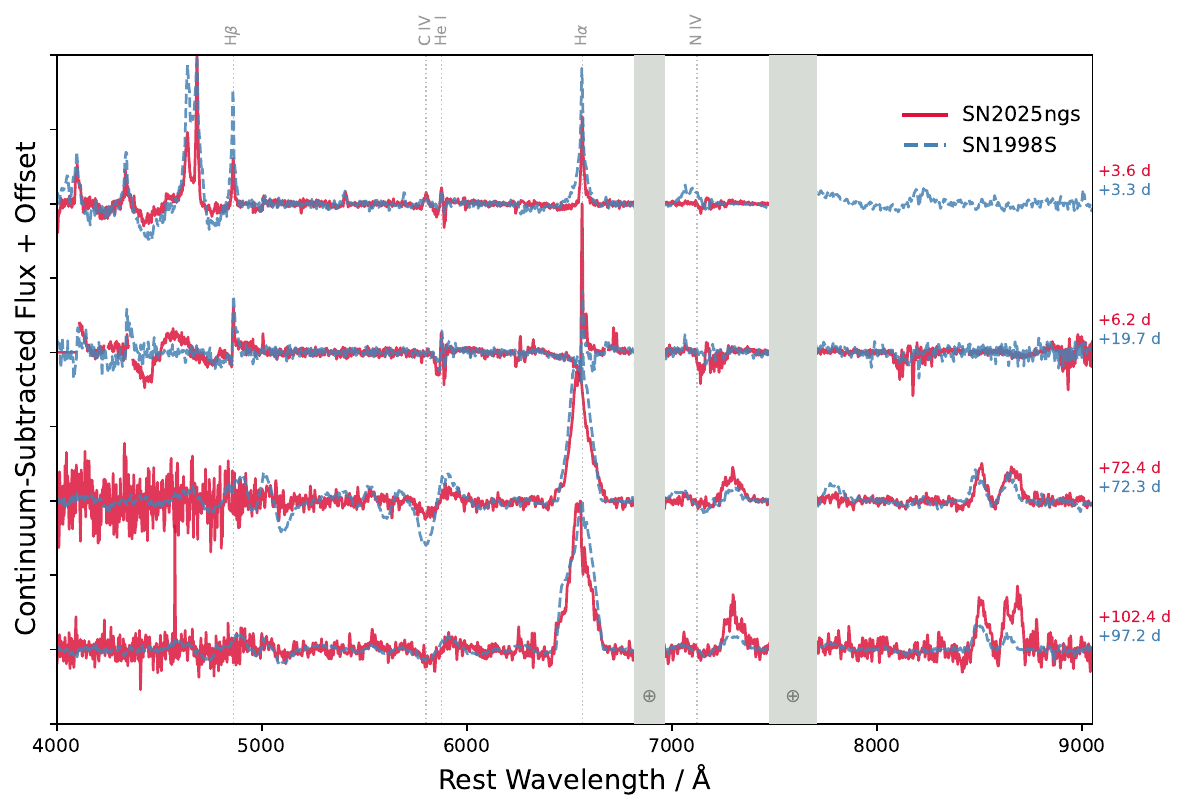}
    \caption{A comparison between the spectra of SN\,2025ngs and SN\,1998S at similar epochs. The SN\,1998S spectra are presented as dashed lines, and SN\,2025ngs is shown as a solid line. We see that these spectra exhibit similar features at the same epochs. The red spectra show the early phase, around 3 days post-explosion for both SNe, exhibiting classic, high-ionization flash features. The middle spectrum exhibits asymmetric line profiles, suggestive of asymmetric CSM. The final epoch, around 2 months post-explosion for both objects shows a broad H$\alpha$ profile, with a possible narrow core, and boxy features, indicative of an exposed shock-front. There is no broad P-Cygni feature on the Balmer profiles at this last epoch, suggestive of interaction.}
    \label{fig:98sspeccomp}
\end{figure*}

After the narrow features fade, H$\alpha$ is left as the most prominent emission feature. At this point, the H$\alpha$ profile starts to become asymmetric, with absorption features starting to form. This evolution is consistent with that seen in similar objects such as SN\,1998S \citep[][]{Leonard_2000, Fassia_2001}, SN\,2024bch \citep[][]{Andrews_2025} and SN\,2024cld \citep[][]{Killestein_2025}. We show the spectroscopic similarity to SN\,1998S in Figure \ref{fig:98sspeccomp}. These features mark the epoch at which the ejecta has emerged from the CSM. At this point, the broader P-Cygni features start to form.

\begin{figure}
    \centering
    \includegraphics[width=0.99\columnwidth]{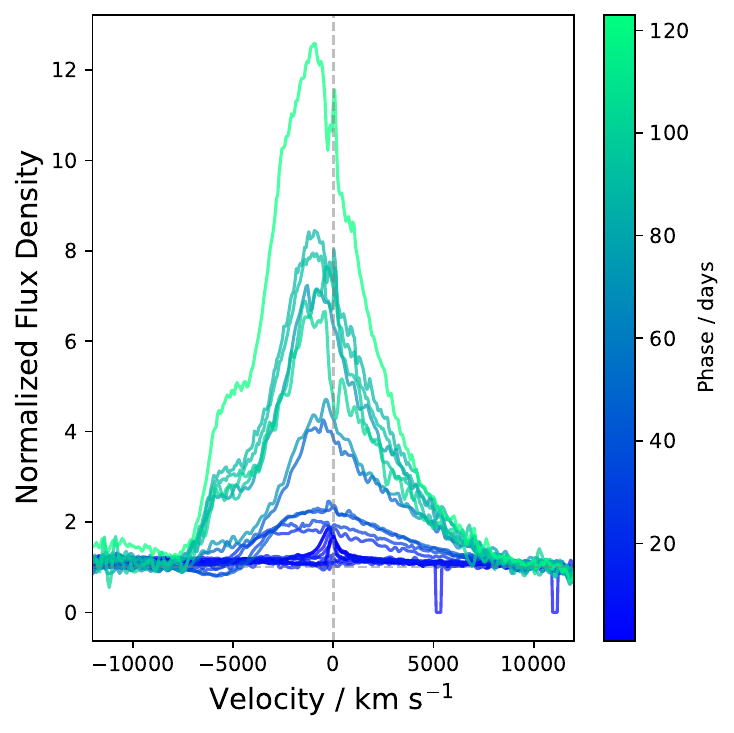}
    \caption{The evolution of the H$\alpha$ profile of SN\,2025ngs. These spectra are normalized to the local continuum and shifted to velocity space. It is apparent that as the transient evolves, the H$\alpha$ peak becomes more blue shifted, with a blue shelf developing. There is a narrow feature centered at zero which may be host contamination. }
    \label{fig:halphaevolve}
\end{figure}

The `standard' SN\,II-like spectrum persists through the short plateau. At the onset of the post-plateau decline, when the outer ejecta becomes more transparent, revealing some of the inner ejecta, the spectrum drastically evolves. The P-Cygni profiles on the Balmer lines subside by around 70 days post-explosion, and the strong H$\alpha$ line develops a more complex profile shape. At this phase, features from the inner ejecta become more apparent such as strong Ca\,II lines, however, this transient is not at the nebular phase at this phase, with no forbidden oxygen features present, or corresponding nebular Balmer features. 

We show the evolution of the H$\alpha$ profile in Figure\,\ref{fig:halphaevolve}. We see that as well as the H$\alpha$ profile becoming more complex, the line also remains strong when compared to the fading continuum. The complex H$\alpha$ feature exhibits a `blue shoulder' emission feature to the blue side of the line core after the absorption features have faded, extending to around $-6,000$\,km\,s$^{-1}$, with this extent not changing throughout the rest of our time-series. This blue shoulder feature is attributed to the shock between an asymmetric (aspherical) CSM and the ejecta \citep[see][]{Leonard_2000,Fassia_2001, Brennan_2023, Killestein_2025}. The H$\alpha$ profile is centered on a narrow feature which may originate from the host, or unshocked CSM, but is likely the former given the asymmetric evolution of the profile. This narrow H$\alpha$ feature is apparent throughout the photospheric phase. Moreover, during this later phase of evolution, the broad H$\alpha$ profiles are blue-shifted. A similar evolution was also seen in our comparison objects, SN\,1998S, and PTF\,11iqb, as well as SN\,2023ixf \citep[e.g.][]{Zheng_2025}, SN\,2024bch \citep[][]{Andrews_2025}, SN\,2024cld \citep[][]{Killestein_2025}, and SN\,2024ggi \citep[][]{Shrestha_2024_24ggi}. The blue-shifted profile may be attributed to dust formation, but, as dust attenuation is wavelength dependent, and we do not see a similar shift in other lines, this feature may indicate asymmetry in the ejecta and/or the CSM, or attenuation by the cold dense shell (the interface between the shock and CSM). The H$\alpha$ profiles at these evolutionary stages may be approximated by a Gaussian decomposition \citep[e.g.][and see Figure\,\ref{fig:append}]{Ransome_2021} in order to estimate velocity components (albeit, omitting the blue shelf). The narrower core of the blue-shifted profile has a FWHM of $\sim1500$\,km\,s$^{-1}$, with a velocity offset of the centroid of this feature at $\sim-1000$\,km\,s$^{-1}$.


In our NIR spectra, we see similar evolution (shown in Figure\,\ref{fig:mmirs}), with broad P-Cygni features on hydrogen lines such as Pa\,$\beta$, which evolve into broad emission features at later times. We again see the strong Ca\,II features develop towards the end of this time series. While there are few clear features in our NIR spectra, we do also see the evolution of Mg\,I features around 1.2\,$\mu$m, as well as He\,I and O\,I, which are common features in CCSN NIR spectra. Notably, in our final IRTF spectrum, there is a possible detection of the first CO overtone, at 95 days post-explosion. This CO overtone at $2.29-2.40$\,$\mu$m is a precursor to dust formation, allowing for rapid cooling and dust condensation. This feature was seen at a similar time in the NIR spectra of SN\,1998S \citep[][]{Fassia_2001}, and at an earlier epoch than SN\,2023ixf \citep[][]{Li_2025, Park_2025, Medler_2025}.

At the end of our spectral time-series (around 60\,--\,120 days post-explosion), these spectra are coincident with the fall of the light curve from the plateau phase. The main features present are the complex H$\alpha$ profile, with the narrow core and persistent blue shelf, along with strong calcium features. These calcium lines are also seen in SN\,1998S, at a similar relative strength at $\sim70$ days. After this point, these features are relatively stronger in SN\,2025ngs, which are typical of earlier nebular emission in CCSNe \citep[e.g.][]{Dessart_2020}. Other nebular emission features, particularly forbidden emission lines such as [O\,I], are not present in our later spectra, apart from weaker [Ca\,II] features. This may be the consequence of the high densities in the SN environment provided by the dense CSM, thus suppressing forbidden line emission. SN\,2025ngs has not completely reached the nebular phase by our last spectrum -- the broad emission lines are persistent and have not transitioned into the complex, boxy profiles seen in traditional nebular spectra. Signatures of CSM interaction and the SN shock are still present.

\subsection{Comparison to Models}\label{sec:comp}

As we have an early-time spectroscopic time series, we are able to compare our data to models of early spectra. Similarly to SNe\,2023ixf \citep[][]{Jacobson-Galan_2023, Bostroem_2023}, 2024bch \citep{Andrews_2025} and 2024ggi \citep[][]{Shrestha_2024_24ggi}, we compare to the spectral models of \citep[][]{Boian_2019}, and \citep[][]{Dessart_2017}. These models vary the abundances, luminosity, mass-loss rates, and CSM densities to represent the early interaction between SN ejecta and CSM.

\begin{figure*}
    \centering
    \includegraphics[width=0.45\textwidth]{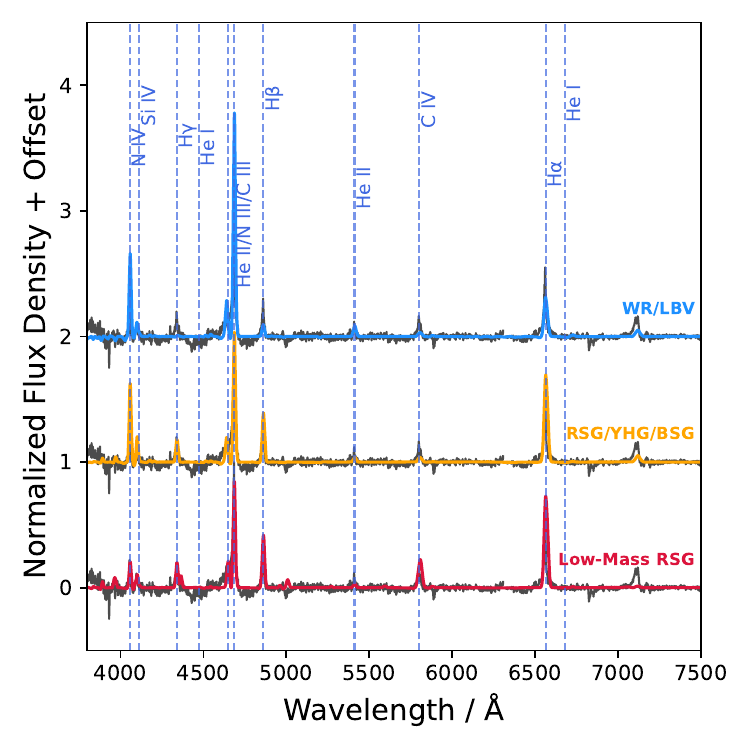}
    \includegraphics[width=0.45\textwidth]{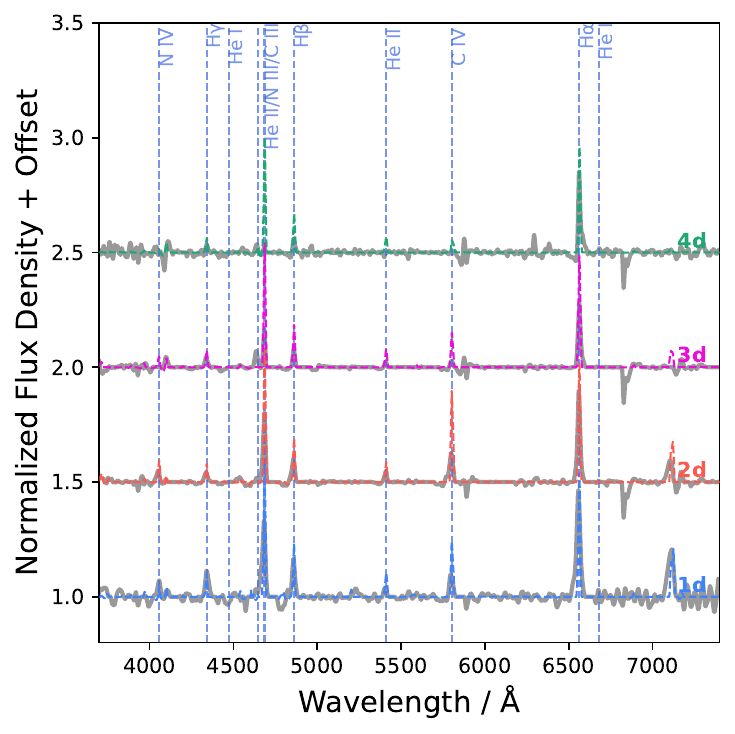}
    \caption{\textit{Left:} Comparisons between the models from \citet{Boian_2019} and the Gemini GMOS spectrum taken a day post-explosion. Here, we compare the $1.5\times10^9$\,L$_\odot$ models, with $\dot{M}=10^{-3}$\,M$_\odot$\,yr$^{-1}$. Each color line represents a different abundance, with the red line being the low mass RSG, the yellow line being consistent with the abundances expected from higher mass RSGs, or YHG/BSGs, and finally, the blue line model represents WR/LBVs. These models cover a single epoch (1 day post-explosion), so we compare our first Gemini/GMOS spectrum to these models. The spectra are normalized with respect to the continuum, and are continuum subtracted for clarity. We also mark the key lines associated with these models, and also with early interaction features generally. \textit{Right:} Comparisons of our early spectra with the corresponding r1w4 \texttt{CMFGEN} models which correspond to a mass-loss rate of 10$^{-3}$\,M$_\odot$\,yr$^{-1}$. We show the models covering the first four days post-explosion as dashed lines. These are of an arbitrary resolution, with a Gaussian smoothing applied for clarity. The black lines show our data at the corresponding epochs. We also mark key lines, particularly those associated with `flash' features. Note that some of these models show narrow P-Cygni absorption features which may not manifest in our spectra due to resolution effects, but are perhaps alluded too through asymmetry in the profiles.}
    \label{fig:speccomp}
\end{figure*}

Firstly, we present the comparison of our earliest spectrum to the models of \citet{Boian_2019} in Figure\,\ref{fig:speccomp} \citep[based on CMFGEN simulated spectra][]{Hillier_1998, Hillier_2012}. The spectral models of \citet[][]{Boian_2019} only extend to around a day post explosion, but vary the abundances in order to represent different progenitor scenarios. Solar abundances replicate what may be expected from a lower mass RSG, enhanced CNO processed material abundances being representative of a more massive RSG, or a more evolved YHG or BSG, and a more helium-rich star such as a Wolf-Rayet or luminous blue variable. These models vary both the SN luminosity and mass-loss rate. As shown in Figure\,\ref{fig:speccomp}, we find that the closest models to our Gemini GMOS data are those with $L_{\mathrm{SN}} = 1.5\times10^9$\,L$_\odot$, and $\dot{M}=10^{-3}$\,M$_\odot$\,yr$^{-1}$, similar to SN\,2023ixf \citep[][]{Jacobson-Galan_2023, Bostroem_2023}, and SN\,2024ggi \citep[][]{Shrestha_2024_24ggi}. No single model matches with our Gemini GMOS spectrum, whether this be the presence of some lines in general, or relative strengths of the emission. The spectrum is, however, mostly consistent with the CNO processed abundances consistent with the more massive RSGs, YHGs and BSGs. There are many similarities with the solar abundance model also, but the model has weaker N\,IV $\lambda7123$ and He\,II $\lambda5411$ lines, and also has lines that are not seen in our data such as O\,III $\lambda5007$, and He\,I $\lambda3998, 4387$. On the other hand, the relative intensities of N\,IV, and C\,IV are a better match at solar abundances, but the He\,II line strengths are a closer match in the CNO processed abundances. As these models only cover the first day spectrum, we can not extrapolate through the evolution of the transient.

In Figure\,\ref{fig:speccomp}, we compare our early spectra to the model suite provided by \citet[][]{Dessart_2017}. In this model grid, simulated spectra are provided for a RSG exploding in CSM with masses up to 0.1\,M$_\odot$, and at radii of up to 10\,$R_*$, where the radius of the star is constant for most models at 501\,$R_\odot$. Via visual inspection of the simulated spectra, we select the \texttt{r1w4} as the closest match to our early time-series. In Figure\,\ref{fig:speccomp}, we trace the first $\sim$four days of spectral evolution. The spectra are normalized and continuum subtracted, and the model spectra are convolved with a Gaussian kernel to approximate the instrumental resolution of our medium dispersion data. The \texttt{r1w4} model represents a RSG with a mass-loss rate of $\sim$10$^{-3}$\,M$_\odot$\,yr$^{-1}$. While this is consistent with the models presented by \citet{Boian_2019}, these models only assume a RSG progenitor. This model was also found to be a good match to SNe\,2022jox \citep[][]{Andrews_2024}, 2023ixf \citep[][]{Bostroem_2023}, 2024bch \citep[][]{Andrews_2025} and 2024ggi \citep[][]{Shrestha_2024_24ggi}.

\subsection{The Circumstellar Medium of SN\,2025ngs} \label{sec:csm}

We have shown that both the early, and late evolution of SN\,2025ngs exhibit features consistent with there being CSM interaction. Indeed, the spectroscopic evolution is congruent with features seen in other objects such as SN\,1998S. As discussed in Section\,\ref{sec:ext}, we have high-resolution early-time spectroscopic observations of SN\,2025ngs. These were taken on 14th and 15th June 2025 using MAROON-X on Gemini-N (as part of the Gemini Large and Long Program, 2024B-LP-112, PI: Sand\,\&\,Andrews) with $R\approx80,000$. Such high resolution, early spectroscopic data are highly informative of the proximate CSM, such as the velocities and geometry \citep[e.g., the case of the nearby SNe\,2023ixf and 2024ggi,][]{Smith_2023_ixf, Shrestha_2024_24ggi,Dickinson_2025}.

 The 14th June H$\alpha$ profile from MAROON-X is shown in Figure\,\ref{fig:maroonx} (red arm). We show the combined spectra from fibers 2 and 3, as fiber 4 may be effected by correlated noise from calibration steps during the reduction process, therefore, holds low SNR data. These spectra reveal a complex H$\alpha$ profile, in addition to the expected electron scattering wings and narrow core. In the spectrum from 14th June, we see that the H$\alpha$ core has a boxy shape, notably with a double peak structure, particularly visible in the red arm data. The blue side of MAROON-X has low sensitivity around H$\alpha$, therefore the SNR is low, and this double peaked feature is not as clear. By the next day, on 15th June, this double peaked feature is no longer apparent, however there may still be a boxy shape/flat top to the line core. The other strong features in these high-resolution data include the quickly evolving carbon lines in Figure\,\ref{fig:earlycomp}, and the interstellar sodium absorption lines that we use to estimate the host extinction.


We show a decomposition of the H$\alpha$ profile from the 14th June spectrum in Figure\,\ref{fig:maroonx}. Firstly, we subtract the continuum from the spectrum using a window around the H$\alpha$ profile. We then fit a series of Gaussian profiles to the data, with a broad profile fitting the electron scattering wings, and then to the narrow component in the line core. We find that the broad component has a FWHM of $\approx$250\,km\,s$^{-1}$. The offset of each peak in the double peaked profile\footnote{Reminiscent of the Tower of Barad-d\^{u}r, with the Eye of Sauron being represented by SN\,2025ngs \citep[][]{Tolkien1954b}, similar to the Eye of Sauron Nebula \citep{Smith_2007_rings}.} is $\approx$20\,km\,s$^{-1}$ either side of the center of the profile. This profile is consistent with a disk-like CSM, possibly with a cavity in the center \citep[e.g.][]{Jerkstrand_2017, Harvey_2020}. The next day, the spectrum has evolved past this clear double-peaked regime. These features may suggest that the early shock/radiation interacts with a confined, dense disk of CSM. Assuming that the disk is initially photo-ionized prior to being swept up by a shock, we estimate the extent of the disk-like CSM being $\sim$10$^{14}$\,cm, which is typical of the CSM associated with early interaction features \citep[e.g.][]{Kochanek_2019, Jacobson-Galan_2024}. Therefore, assuming that the CSM has not been significantly (de)accelerated, the CSM disk was formed 10\,--\,20 years ago (assuming CSM velocity is $\approx$20\,km\,s$^{-1}$ for this region of the CSM). This is the first example of a double horned H$\alpha$ profile, therefore, strobg evidence for a SN shocking a confined, disk-like structure.

Aside from our high resolution data, our spectral time-series presented in Figure\,\ref{fig:spec} shows that by day 72 (with a 24 day gap from the previous spectrum), the P-Cygni feature has disappeared, yielding to a blue shoulder on the H$\alpha$ profile. This blue shoulder is a signature of newly re-revealed CSM interaction, exposing the shock front. Converse to interaction with some higher density shell, or clump of CSM, which would be accompanied by a bump in the light curve (as $L\propto \rho^2$), and also narrow features from the subsequent recombination of photoionized, unshocked CSM, this CSM interaction region was within the (pseudo-)photosphere of the SN. As the ejecta expands, and becomes less dense (i.e. less optically thick as the ejecta recombines), the photosphere recedes, revealing the edge of the shock front near the cold dense shell (the discontinuity between the forward and reverse shock formed by CSM interaction). This feature is seen in many SNe\,IIn \citep[e.g.][albeit somewhat blended with electron scattering wings due to much denser, more massive CSM]{Fransson_2014_10jl, Taddia_2020, Fransson_2022,Brennan_2023} and is also seen in other intermediate interacting SNe, notably SN\,1998S \citep{Fassia_2001}, PTF\,11iqb \citep{Smith_2015}, and SN\,2024cld \citep{Killestein_2025}.


\subsection{Comparisons to Similar Supernovae}

SN\,2025ngs is the latest member of intermediate interacting SNe that (spectroscopically) resemble SN\,1998S (sometimes known as 98S-like SNe). Spectroscopically, the early evolution is typical of other SNe that exhibit flash features, including SN\,1998S. Here, we discuss the comparison between this SN and similar transients, and groups of SNe that share some properties with SN\,2025ngs.

\subsubsection{Flash Supernovae}

 We have shown in Section\,\ref{sec:data} that the early epochs of SN\,2025ngs are similar to that of other SNe that show so-cakked flash features. This is true even for objects that go on to evolve as typical SNe\,II. We show comparisons to the nearby, well studied SNe 2023ixf and 2024ggi in Figures\,\ref{fig:col}, \ref{fig:spec_comps}, and \ref{fig:earlycomp}. Firstly, the color evolution of SN\,2025ngs is similar to both SNe\,2023ixf and 2024ggi (being most similar to SN\,2024ggi). The later color evolution has a similar general evolution to SN\,1998S, but is redder. The flash features are similar between these comparison objects at early times. However, as the flash features in SNe\,2023ixf and 2024ggi disappear within a week, they do not form narrow P-Cygni features \citep[][]{Bostroem_2023, Shrestha_2024_24ggi}. This is contrasted by SN\,1998S and SN\,2025ngs, which do form narrow P-Cygni features. This can be seen in the spectra around days 4\,--\,15 for SN\,2025ngs. These features suggest that the CSM in SN\,2025ngs is more extended than in other flash SNe, similarly to what is inferred in SN\,1998S. After $\sim$20 days, broad P-Cygni features form, initially manifesting as a relatively shallow feature. At later times, SN\,2025ngs loses the broad P-Cygni features and a complex H$\alpha$ profile evolves. This is contrasted to SN\,2023ixf and SN\,2024ggi, which continue to evolve as relatively normal SNe\,IIP with broad P-Cygni profiles with clear, deep absorption. Moreover, when the early high resolution spectra are compared at a similar epoch, SN\,2023ixf and SN\,2024ggi are almost identical with a Lorentzian H$\alpha$ profile with a helium line on the blue side of hydrogen. The high-resolution, early spectrum of SN\,2025ngs, however, diverges from these features, while the H$\alpha$ profile has a general Lorentzian shape expected from CSM interaction, the peak of the profile exhibits the double horned morphology, and does not exhibit helium. This may suggest that the progenitor system of SN\,2025ngs is distinct from SNe\,2023ixf, and 2024ggi.

While SN\,2025ngs shows the classic flash lines seen in many SNe\,II \citep[][]{Bruch_2023}, the key differences, such as the short plateau, prolonged early narrow features, and the general spectral evolution points towards a different picture. The CSM interaction in SN\,2025ngs is stronger than in most SNe with early interaction signatures. Some other interacting SNe exhibit flash features (e.g. SN\,1998S and 2024cld), but the strongly interacting SNe\,IIn do not commonly exhibit the high ionization lines associated with the shock interaction of confined CSM containing CNO products mixed in the material on the surface of the progenitor \citep[e.g. see][]{Khazov_2016, Kochanek_2019}. This is due to the much denser, massive hydrogen-rich CSM around SNe\,IIn. In the case of SN\,2025ngs, the early spectral evolution suggests that the CSM is more extended than in SNe like SN\,2023ixf or SN\,2024ggi, while not being as extreme as in SNe\,IIn.


\subsubsection{SN\,1998S and PTF\,11iqb}

Both SNe\,1998S and PTF\,11iqb may be considered as transitional interacting SNe, which bridge the gap between normal SNe\,II, and the strongly interacting SNe\,IIn \citep[e.g.][]{Smith_2015, Shivvers_2015}. SN\,1998S was also one of the earliest examples of a SN exhibiting flash features \citep[][]{Leonard_2000,Fassia_2001,Shivvers_2015}. We have shown that the spectroscopic evolution of SN\,2025ngs is remarkably similar to that of SN\,1998S (see Figure\,\ref{fig:speccomp}). This is in contrast with the photometric evolution of SN\,1998S, which peaks over a magnitude brighter than SN\,2025ngs, and also has a SN\,IIL-like light curve \citep[][]{Fassia_2000}. SN\,1998S similarly has a complex H$\alpha$ profile, with no P-Cygni profile being exhibited, suggestive of back-heating of the ejecta due to interaction. SN\,1998S also showed interaction for many years post-explosion, with a complex multi-horned nebular profile on the H$\alpha$ line \citep[][]{Leonard_2000,Mauerhan_2012}, indicative of a complex clumpy CSM, or a disk/ring-like CSM structure. Moreover, spectropolarimetry of SN\,1998S reveals persistent polarization, further suggesting an aspherical CSM \citep[][]{Leonard_2000}. The spectroscopic similarity along with the stark photometric differences demonstrate that the 98S-like SNe may have diverse origins. Conversely, as both SNe exhibit evidence for aspherical, possibly disk-like CSM, the viewing angle may also play a role in the morphology of the light curve \citep[][]{Suzuki_2019}, with more linear declines being attributed to 'face-on' viewing angles, and more plateau-like features from a more edge-on angle.

PTF\,11iqb also exhibited flash lines in the early spectrum, and evolved similarly to both SN\,1998S and SN\,2025ngs \citep[][]{Smith_2015}. \citet{Smith_2015} note that compared to SN\,1998S, the early interaction features fade quickly, and reappear at later times somewhat stronger. The complex H$\alpha$ line that is revealed to evolve almost identically to what is seen in SN\,1998S (and SN\,2025ngs), suggesting an aspherical CSM. Those authors interpret the spectral evolution of PTF\,11iqb as the SN ejecta engulfing the CSM interaction, obscuring the narrow line region. Then the interaction features are revealed again as the SN photosphere recedes. We do not have nebular phase spectra of SN\,2025ngs as of time of writing. Such observations, however, may also reveal complex line profiles, revealing ongoing interaction. 

There are a number of other SNe inferred to have a circumstellar disk-like CSM structure. These are mostly the strongly interacting SNe\,IIn. \citet{Bilinski_2024} presents a sample of SNe\,IIn that exhibit persistent polarization, consistent with persistent aspherical, disk-like CSM geometries \citep[see also][]{Bilinski18}. Additionally, other SNe\,II or transitional SNe\,II have inferred disk-like CSM similar to SN\,1998S, PTF\,11iqb, and SN\,2025ngs, such as SN\,2023ixf \citep{Vasylyev_2026}, and SN\,2024cld \citep{Killestein_2025}.

\subsubsection{Short-plateau Supernovae}

Short-plateau SNe have a light curve plateau significantly shorter in duration than the typical SN\,IIP plateau phase of $\sim100$\,days. The plateau duration of SN\,2025ngs is $\sim70$\,days, similar to the sample of short-plateau SNe 2006Y, 2006ai, and 2016egz presented by \citet{Hiramatsu_2021}, as well as SN\,2018gj \citep[][]{Teja_2023_18gj}, SN\,2020jfo \citep[][]{Sollerman_2021, Teja_2022}, and SN\,2023ufx \citep{Ravi_2025}. SN\,2025ngs is somewhat less luminous than most of the short-plateau sample, with SN\,2020jfo being at a similar luminosity, and SN\,2018gj being somewhat fainter. 

Not all short-plateau SNe exhibit narrow line CSM interaction signatures \citep[e.g. SN\,2020jfo][]{Sollerman_2021}, but show early bumps or `shelves' in the spectra. For example, model fits to SN\,2018gj and SN\,2020jfo require CSM interaction in order to explain the early-time light curves \citep[][]{Teja_2022,Teja_2023_18gj}. SNe\,2020amv and 2023ufx do, however, show clear, early narrow features in their spectra \citep[][]{Sollerman_2021}. These differences, along with what is seen in SN\,2025ngs, may indicate that the observational properties of these short-plateau objects may be influenced by a diversity in progenitor systems, CSM geometry, viewing angle, or a combination of such effects.

\subsubsection{Short-plateau vs Type IIL Supernovae}

SN\,2025ngs is photometrically similar to the short-plateau SNe, and spectroscopically similar to SN\,1998S and related transients. Both of these groups may have semi-stripped progenitors, retaining enough hydrogen to have hydrogen-rich spectra, but not enough to produce a long plateau. \citet{Hiramatsu_2021} presents a set of light curve models demonstrating the effect of different envelope masses on light curve shapes (and the relation to different classes). Short-plateau SNe have a larger hydrogen envelope than the more stripped SNe\,IIb, and a less massive envelope than SNe\,IIL. The pseudobolometric light curve of SN\,2025ngs is consistent with the higher envelope mass short plateau SNe presented by \citet{Hiramatsu_2021}, and is also similar to the lower envelope IIL models. These models have an envelope mass of M$_{\mathrm{H}_{\mathrm{env}}}\approx2\,$M$_\odot$. The inferred envelope mass from our shock-cooling fits in section \ref{sec:shock} suggests that SN\,2025ngs has an envelope mass ranging from the upper end of the short-plateau SNe to the lower end for SNe\,IIL. SN\,2025ngs may, therefore exemplify a continuum between these subclasses. 



\section{The Progenitor of SN\,2025ngs}  \label{sec:prog}

The observed progenitors of SNe\,IIP are RSGs \citep[][]{Smartt_Review}. The progenitors for SNe\,IIL may be more stripped, hence a smaller hydrogen envelope, resulting in the absence of a long recombination powered plateau phase. Such progenitors may include YHGs, or more massive RSGs \citep[e.g.][]{Elias-Rosa_2010, Fraser_2010, Elias-Rosa_2011}. The progenitor landscape of SNe\,IIn is more complicated, with recovered progenitors being associated with massive LBVs \citep[e.g.][]{Gal-Yam_2007, Boian_2018}, but the local environments are inconsistent with young stellar populations and massive stars, suggesting multiple progenitor routes \citep[e.g.][]{hab14, Galbany_2018, Ransome_2022}. SN\,2025ngs is too distant and embedded in the host galaxy to recover a pre-explosion image of the progenitor. In any case, the progenitor of SN\,2025ngs must retain at least some hydrogen envelope in order to exhibit strong hydrogen spectral features, and to power the plateau phase. The progenitor must also suffer from a mass loss event(s) several orders of magnitude stronger than the massive winds of RSGs \citep[][]{Bruch_2023, Hinds_2025}, and also form a disk, or ring like CSM. 



The mass loss of SN progenitors is also strongly governed by binarity. Mass loss may occur via mass transfer during Roche-lobe overflow, common envelope evolution, collisions and mergers. Indeed, SNe\,IIb and Ib/c are thought to have binary progenitor systems due to the mass stripping required \citep[][]{Woosley_1994, Maund_2004,Smartt_Review,Modjaz_2011, Eldridge_2014, Maund_2016, Kilpatrick_2021, Zapartas_2026}. The high mass-loss rates inferred for SNe\,IIn, and the mass-loss rates shortly before some SNe\,II that exhibit early interaction features have also invoked binaries as a possible route to attain such high mass loss \citep[e.g.][]{SmithArnett_2014,Dickinson_2024, Dukiya_2024,Smith_2024,Ercolino_2024, Ransome_2024}. Moreover, the light curves of short plateau SNe can be reproduced after binary evolution of lower mass RSGs using \texttt{BPASS} \citep[][]{Eldridge_2018}, and there may be a continuum between the semi-stripped SNe\,IIb, and short-plateau SNe \citep[][]{Hiramatsu_2021,Farah_2026}. Indeed, the short-plateau SN\,2018gj has a RSG-like progenitor within an old stellar population, and population synthesis models favor a close-binary system to produce such a progenitor in an isolated location \citep[][]{Niu_2026_18gj}. 

The progenitor of SN\,2025ngs was surrounded by a disk of CSM, as we show in our high-resolution spectroscopy. There are Galactic analogs of such a progenitor scenario, for example, the Fried Egg Nebula which surrounds Hen\,3-1379, comprised of multiple detached shells from apparent episodic mass loss \citep[e.g.][]{Lagadec_2011, Koumpia_2020}, and such structure is observed around the RSG that was possibly observed to evolve into a YHG, WOH\,G64 \citep{Munoz-Sanchez_2024, Ohnaka_2024}, and may also be in a binary system with a hot, accreting companion. However, more recent observations suggest that WOH\,G64 is still a RSG \citep[][]{vanLoon_2026}. Moreover, the mass-loss rate of WOH\,G64, and possibly similar stars are of order a few $10^{-4}$\,M$_\odot$\,yr$^{-1}$, an order of magnitude below our estimates from spectral model matching. Close binary interactions, mergers, or common envelope evolution is a natural route to the complex CSM structures observed around Galactic analogs of these SN progenitors, even if no companion is observed, as many massive stars will merge with their companion(s) prior to their deaths \citep[][]{deMink_2014}.

The inferred nickel mass of SNe may inform on the progenitor system. Similar, luminous, short-plateau SNe\,II have a range of inferred nickel masses. For example SNe\,1998S, 2006Y, 2006ai, 2016egz and 2023ufx, all have high inferred nickel masses \citep[][]{Fassia_2001, Hiramatsu_2021, Ravi_2025}, higher than the median (around 0.03\,M$_\odot$) from the samples presented by \citet{Anderson_2014} and \citet{Valenti_2016}. Such high nickel masses (M$_{^{56}\rm Ni}>0.1$\,M$_\odot$) may suggest a massive progenitor star with masses exceeding 17\,M$_\odot$ being suggested, although, other short-plateau SNe\,II, SNe\,2018gj and 2020jfo have typical SN\,II inferred nickel masses, and may have lower mass progenitors ($\approx12$\,M$_\odot$). Notably the SN\,IIn-P subclass that exhibits spectroscopic evolution consistent with SNe\,IIn, and a SN\,IIP-like light curve are also inferred to have low nickel masses \citep[e.g. SN\,1994W, SN\,2011ht, and SN\,2020pvb][]{Sollerman_1998, Mauerhan_2013_11ht,Elias-Rosa_2024}. Furthermore, nebular spectra models, and more modest luminosities, may be associated with a lower mass RSG \citep[e.g.][]{Sollerman_2021}. Our upper limit on the nickel mass of 0.007\,M$_\odot$ may therefore indicate a lower mass progenitor for SN\,2025ngs.

For SN\,2025ngs, we find a high mass-loss rate of around 10$^{-3}$\,M$_\odot$\,yr$^{-1}$ from our spectral model matching presented in Section\,\ref{sec:comp}. We also find abundances similar to a RSG or YSG in the early spectra. We can not make a distinction between the high mass RSG/YSG scenario and the more modest RSG scenario due to the coarseness of the spectral models used, meaning that some lines/intensity ratios are a better match in one model than the other, and vice versa. Due to the inclination of the host, high extinction, and how embedded SN\,2025ngs is within the host, it is not possible to make confident measurements of the local environment in order to estimate the local stellar population properties. This being said, the nickel mass is likely small, so the progenitor mass may be modest. In either case, SN\.2025ngs-like transients are consistent with having binary progenitor systems, where at least the star that exploded was a supergiant.

\section{The Host, NGC\,5961} \label{sec:host}

The host of SN\,2025ngs, NGC\,5961 is a high surface-brightness galaxy towards the Hercules supercluster with a Hubble class of Sb \citep{Tarenghi_1994}. In this section, we will explore the properties of the host by fitting stellar population models to the galaxy photometry and compare to large host samples.


\subsection{\texttt{FrankenBlast} Analysis}

In order to explore the properties of the host galaxy of SN\,2025ngs, we use \texttt{FrankenBlast} \citep[][]{Nugesnt_2025}. \texttt{FrankenBlast} is a rapid, SBI++ \citep{Wang_2023} implementation of \texttt{Blast} \citep[][]{Jones2024}. \texttt{FrankenBlast} firstly utilizes \texttt{Pr{\"o}st} \citep{Prost} for automated, probabilistic host association of transients. Then, photometry is performed on these hosts using public survey data (e.g. Pan-STARRS, DECam, and 2MASS). The resultant host photometry (i.e. the resultant SED) is then fit to stellar population models using \texttt{Prospector} \citep[][]{jlc+2021}. From these stellar population model fits, we can then infer stellar population properties.

\begin{figure*}
    \centering
    \includegraphics[width=0.99\textwidth]{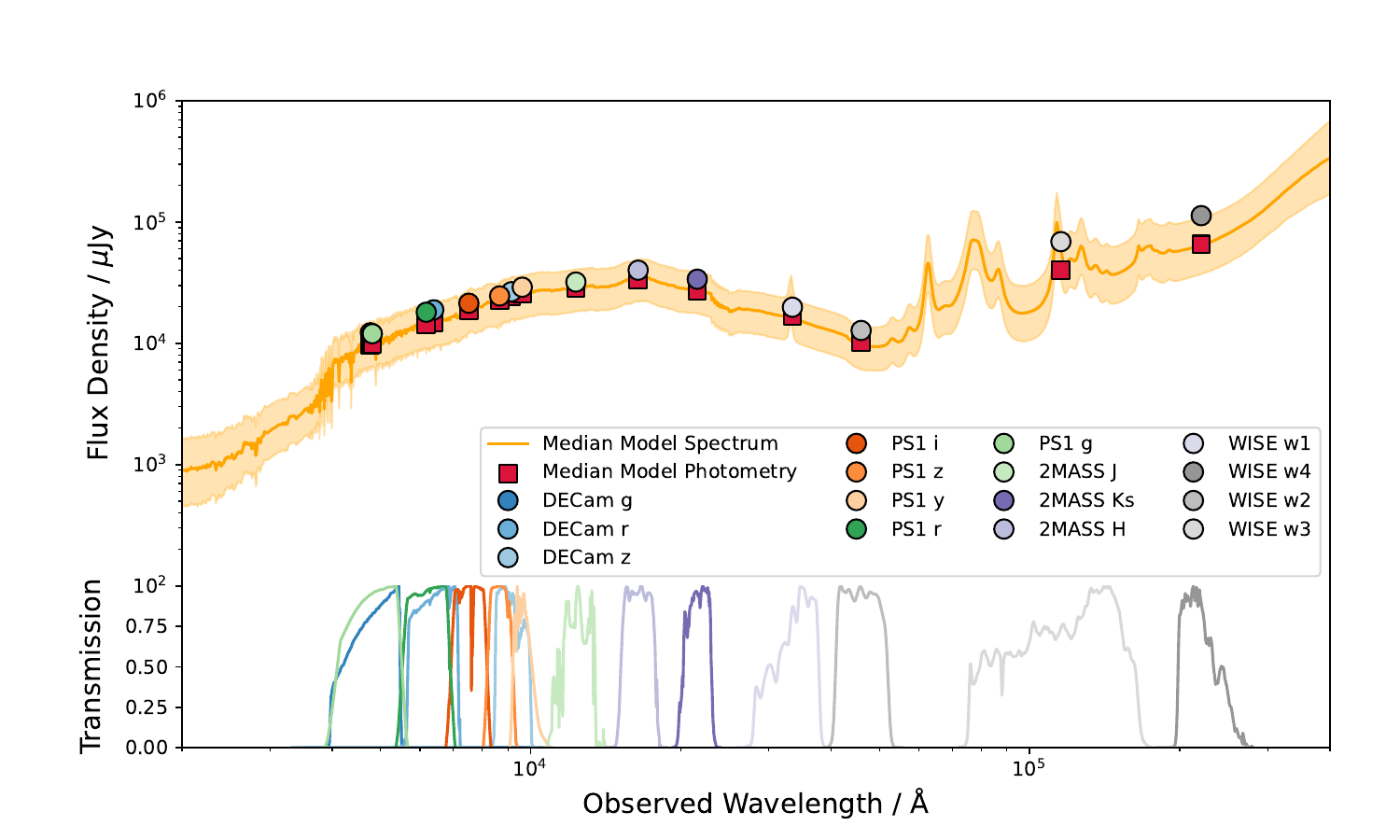}
    \caption{The best-fit SED for the host of SN\,2025ngs, NGC\,5961 from \texttt{FrankenBlast}. The data used in the construction of the host SED include observations from DECam $grz$-bands, Pan-STARRS $grizy$-bands, 2MASS $JKsH$-bands, and WISE $w1, w2, w3, w4$-bands. Shown as an orange line is the best-fit model spectrum, and the shaded region represents the 1$\sigma$ uncertainty. Red squares show the best fit photometry for each filter. The transmission functions of each filter are also plotted below the SED. }
    \label{fig:blast}
\end{figure*}

We show the \texttt{FrankenBlast} SED fits to stellar population models in Figure\,\ref{fig:blast}. Our host analysis suggests that while NGC\,5961 is not a dwarf galaxy, it is relatively low mass, with log($M_*/M_\odot$)\,=\,9.37$^{+0.19}_{-0.25}$. The median age of the stellar populations is 6.82$^{+0.19}_{-0.25}$\,Gyr. The metallicity of the host may be subsolar with log($Z_*/Z_\odot$)\,=\,$-$0.18$^{+0.14}_{-0.20}$, however solar metallicities are consistent within the spread of the posterior distribution. A clean host spectrum is required to get a more constraining value of the metallicities as narrow host lines are used as diagnostics. The star-formation rate (SFR) of NGC\,5961 is 0.25$^{+0.37}_{-0.17}$\,M$_\odot$\,yr$^{-1}$, with a corresponding specific star-formation rate (SSFR) being -9.97$^{+0.64}_{-0.69}$\,yr$^{-1}$. This SSFR is higher than that of the Milky Way \citep[a few 10$^{-11}$\,yr$^{-1}$, e.g.][]{Licquia_2015}, suggesting a strongly star-forming galaxy, consistent with the Hubble type and color. We summarize these properties in Table\,\ref{tab:host}. To summarize, NGC\,5961 is a less massive spiral galaxy which is strongly star forming, with sub-to-solar metallicity. We will now compare these parameters to broader populations. 

\begin{figure*}
    \centering
    \includegraphics[width=0.99\columnwidth]{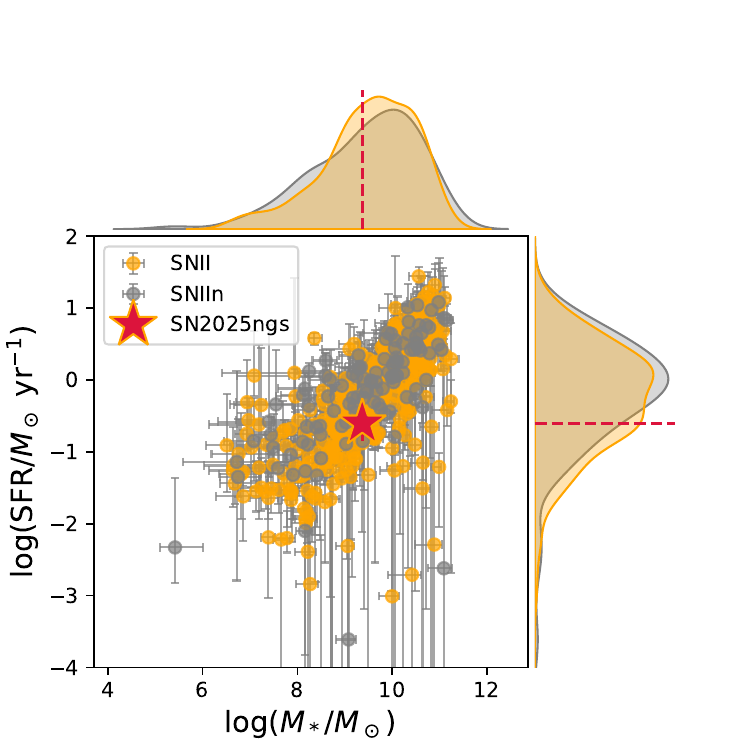}
    \includegraphics[width=0.99\columnwidth]{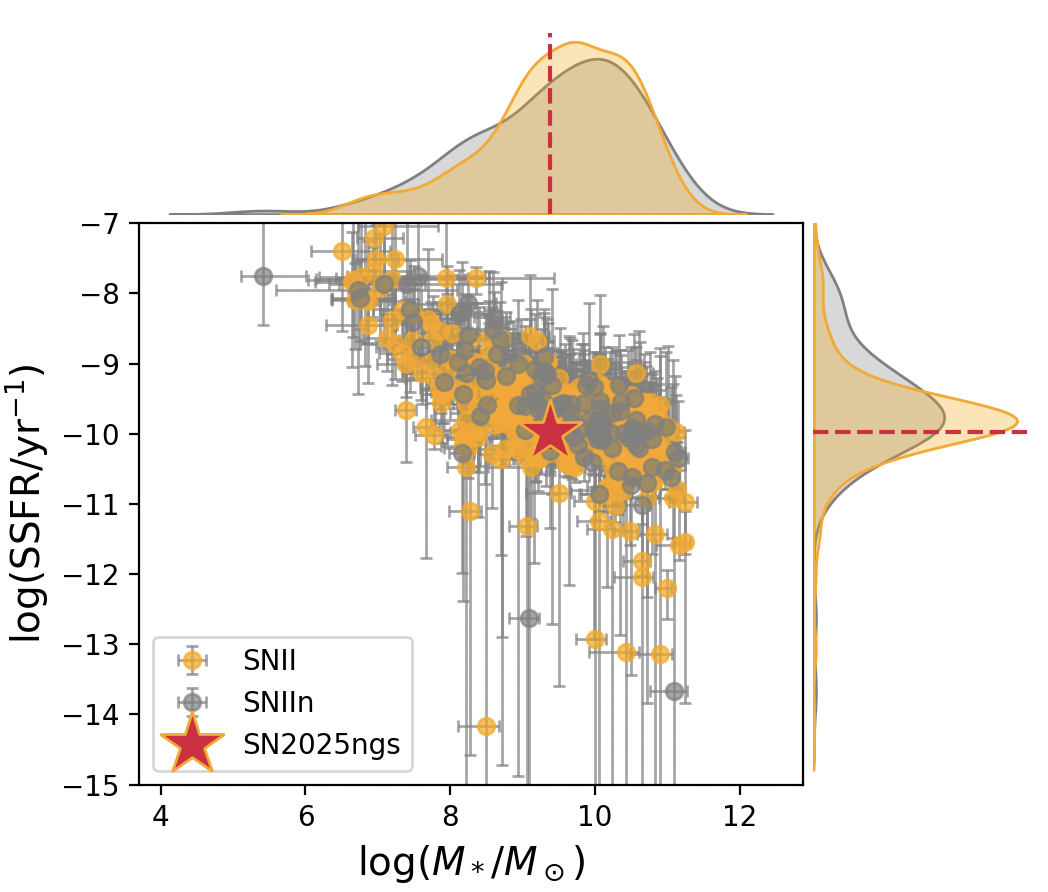}
    \caption{\textit{Left:} The parameter pair distribution of the star formation rate, and stellar masses of the hosts analyzed by \citet{Schulze_2021}. The red star denotes the location of the host of SN\,2025ngs in this distribution. We compare NGC\,5961 to both the SN\,II and SNe\,IIn host distributions. We find that NGC\,5961 is not an outlier in either distribution, is close to the peaks of the stellar mass distribution for both comparison populations. The star formation rate of NGC\,5961 is slightly lower than the peak of the comparison sets. \textit{Right:} Same as left, but comparing against the specific star formation rate. We note that the specific star formation rate of NGC\,5961 is typical when compared to both SN\,II and SN\,IIn hosts, albeit with the hosts of SN\,IIn having a higher specific star formation rate tail. }
    \label{fig:mass_v_sfr}
\end{figure*}

We compare the global properties of NGC\,5961 to the sample presented by \citet{Schulze_2021}. Those authors analyze a sample of 888 SN host galaxies. The SNe in these galaxies range from stripped-envelope SNe to superluminous SNe\,IIn. Similarly, they fit SEDs from the host photometry of their sample. In Figure\,\ref{fig:mass_v_sfr}, we compare the distribution of the stellar masses and the SFR of the sample of SNe\,II and SNe\,IIn from \citet{Schulze_2021} with our values for SN\,2025ngs. We also show the distributions of each axis in Figure\,\ref{fig:mass_v_sfr}, marking the location of SN\,2025ngs within these distributions. NGC\,5961 is not an outlier in either distribution, with the host being within the main `cluster' of the parameter pair distribution. The peaks of the mass distribution for both the SNe\,IIn and SNe\,II are around 10$^{11}$\,M$_\odot$, while the stellar mass of NGC\,5961 is a few $\approx$\,10$^9$\,M$_\odot$. The SFR of the host of SN\,2025ngs is somewhat lower than the majority of the hosts for both SNe\,II and SNe\,IIn. However, a more comparable measure is the specific SFR (SSFR, the SFR normalized by the host stellar mass). We show the SSFR distribution in Figure\,\ref{fig:mass_v_sfr}. Here, we see that the SSFR is consistent with the SSFR distribution peak of both SNe\,II and SNe\,IIn.


\begin{table}
    \centering
    \begin{tabular}{l|c}
        \hline
         Parameter & Posterior Median and Spread\\
         \hline
         log($M_*/M_\odot$)& 9.37$^{+0.19}_{-0.25}$ \\
         log($Z_*/Z_\odot$)& -0.18$^{+0.14}_{-0.20}$\\
         Age / Gyr& 6.82$^{+0.19}_{-0.25}$\\
         A$_V$ / mag& 0.69$^{+0.37}_{-0.28}$\\
         SFR / M$_\odot$\,yr$^{-1}$& 0.25$^{+0.37}_{-0.17}$\\
         log(SSFR / yr$^{-1}$)& -9.97$^{+0.64}_{-0.69}$\\
    \end{tabular}
    \caption{Summary of the properties of NGC\,5961, the host of SN\,2025ngs. These host parameters were inferred using \texttt{FrankenBlast}. We report the medians of each posterior distribution and the 16th and 84th percentile spread (i.e. 1$\sigma$) of the posterior distributions.}
    \label{tab:host}
\end{table}

\section{Conclusion and Summary} \label{sec:conc}

In this paper, we present the SN\,II, SN\,2025ngs through NUV-to-NIR photometry and spectroscopy. We find that SN\,2025ngs is a short-plateau SN with signatures of CSM interaction in both the early time spectroscopy and light curve, and in the late-time evolution, revealing enduring interaction. SN\,2025ngs closely resembles other SNe\,II with properties in between that of normal SNe\,II and SNe\,IIn such as SN\,1998S, with early features similar to other SNe exhibiting flash features. Our findings can be summarized thusly:

\begin{enumerate}
    \item SN\,2025ngs was discovered on 2025-06-12 by ATLAS in NGC\,5961. SN\,2025ngs was rapidly followed up photometrically and spectroscopically.
    \item Early spectra reveal SN\,2025ngs to be a SN\,II with early flash features consistent with interaction with confined dense CSM. We find a redshift of $z=0.0065$. Using the Na\,ID doublet seen in our early high resolution spectrum, we find a host reddening of 0.152\,mag, with a total extinction of $A_V=0.549$\,mag, consistent with fits to the host SED.
    \item SN\,2025ngs peaked at $M_R=-17.9$\,mag, relatively luminous for SNe\,II. SN\,2025ngs evolved to be a short-plateau SN, with a plateau duration of $\sim70$\,days, on the upper end of the short-plateau range \citep[e.g.][]{Hiramatsu_2021}. SN\,2025ngs also had a rapid decay ($s_1$), similar to other short-plateau SNe such as SN\,2006Y and SN\,2006ai. The color evolution of SN\,2025ngs is similar to SN\,2024ggi, being somewhat redder than SN\,1998S.
    \item Fits to the early light curve shows the explosion time to be MJD\,60837.8. The post-plateau tail did not settle to a slope consistent with radioactive decay, possibly indicating late-time interaction. A low nickel mass upper limit of 0.007\,M$_\odot$ was inferred from this late-time data.
    \item The spectroscopic evolution exhibits typical flash lines at early times, evolving into a fairly standard SN\,II spectrum with broad emission features and P-Cygni profiles. The spectrum evolved quickly, and by day 70, as the plateau starts to decline, the P-Cygni absorption subsides, and a complex H$\alpha$ profile develops. This is interpreted as interaction regions being revealed by the receding photosphere, similar to other transients such as SN\,1998S, PTF\,11iqb, and SN\,2024cld. We also show flash features in the NIR, such features are seldom observed.
    \item While generally, the early spectra are typical of early flash interaction SNe, there is a unique feature seen in our high resolution early spectra that shows that SN\,2025ngs has distinct properties from the handful of other objects with similar data.  In particular, the first Maroon-X spectrum displays a double-horned, symmetric profile with peaks at $\pm20$\,km\,s$^{-1}$ from the central emission. This feature is evidence of a ring- or disk-like CSM structure, whichfaded by the next day, providing a lower limit on the extent of this structure of a few 10$^{14}$\,cm.
    \item We fit the early photometry to shock-cooling models. We find that the early-time fits do not capture the rise, similar to that seen in other flash SNe such as SN\,2024ggi \citep[][]{Shrestha_2024_24ggi}. These poor fits are typical when extra luminosity is introduced to the early evolution through CSM interaction.
    \item While photometrically diverse, there are spectroscopically similar SNe to SN\,2025ngs. Notably, SN\,1998S, which while more luminous, with a linear decline, closely resembles SN\,2025ngs throughout its evolution. 
    \item When compared to the \texttt{CMFGEN} models presented by \citet{Dessart_2017} and \citet{Boian_2018}, we find that the early spectra are best matched with models that have a high mass-loss rate (\texttt{r1w4}) of $\sim10^{-3}$\,M$_\odot$\,yr$^{-1}$. We can not make a confident distinction between the models with abundances that represent lower mass RSGs or higher mass RSGs/YHGs, but our spectra are a good match to both scenarios.
    \item The short-plateau, and complex mass-loss history, along with our spectra point to the progenitor of SN\,2025ngs being a supergiant, which was partially stripped in an interacting binary system. Our small upper limit on the nickel mass, however, may suggest a lower mass progenitor.
    \item The light curve shape, inferred shock cooling parameters, and the light curve models presented by \citet{Hiramatsu_2021} suggest that SN\,2025ngs is an intermediate object between short-plateau SNe and SNe\,IIL. This indicates a possible continuum between these classes. 
    \item The host of SN\,2025ngs is NGC\,5961. We fit the host SED with \texttt{FrankenBlast} in order to estimate the global host properties. We find that NGC\,5961 is a fairly low-mass galaxy, with a mass a few 10$^9$\,M$_\odot$ and roughly solar metallicity and a high star-formation rate. When compared with a sample of SN\,II and SN\,IIn, we find that the mass is slightly lower than the peak of the mass distribution for either population and a typical SSFR.
    
\end{enumerate}

SN\,2025ngs is the latest member of the short-plateau SN family, and is also a spectroscopic mimic of SN\,1998S. This transient uniquely highlights the complex circumstellar environments of interacting supernovae, exhibiting evidence for a ring-like CSM. SN\,2025ngs is an example of a transient bridging the gap between normal  SNe\,II that show flash features, and the strongly interacting SNe\,IIn. The growing number of these transitional, or hybrid interacting SNe may be an indication of a continuum of interacting transients, and therefore pre-SN mass loss. With the commencement of the Legacy Survey of Space and Time at the Vera C. Rubin Observatory, the gap may be filled in the coming discovery deluge.

\begin{acknowledgments}

We thank Jan Eldridge and Alexander Heger for their helpful discussions. We thank Andreas Seifarht and the MAROON-X team for their insight into our high-resolution data products. Time-domain research by the University of Arizona
team and D.J.S. is supported by National Science Foundation
(NSF) grants 2308181, 2407566, and 2432036. This
work makes use of data from the Las Cumbres Observatory global
telescope network, which is supported by NSF grant AST-
-2308113. S.V. and the UCDavis time-domain
research team acknowledge support by NSF grants AST-2407565.
J.E.A. and T.R.G are supported by the international Gemini
Observatory, a program of NSF's NOIRLab, which is managed by
the Association of Universities for Research in Astronomy
(AURA) under a cooperative agreement with the National Science
Foundation, on behalf of the Gemini partnership of Argentina,
Brazil, Canada, Chile, the Republic of Korea, and the United
States of America. K.A.B. is supported by an LSST-DA Catalyst
Fellowship; this publication was thus made possible through the
support of grant 62192 from the John Templeton Foundation to
LSST-DA. S.H.P. and S.-C.Y. are supported by the National
Research Foundation of Korea (NRF RS-2024-00356267). N.F. acknowledges support from the National Science Foundation Graduate Research Fellowship Program under Grant No. DGE-2137419. Supernova research at Rutgers University is supported in part by NSF grant AST-2407567.
This work has made use of data from the Asteroid
Terrestrial-impact Last Alert System (ATLAS) project. The
Asteroid Terrestrial-impact Last Alert System (ATLAS)
project is primarily funded to search for near-Earth asteroids
through NASA grants NN12AR55G, 80NSSC18K0284,
and 80NSSC18K1575; by-products of the NEO search
include images and catalogs from the survey area. This work
was partially funded by the Kepler/K2 grant J1944/
80NSSC19K0112, HST GO-15889, and the STFC grants
ST/T000198/1 and ST/S006109/1. The ATLAS science
products were made possible through the contributions of the
University of Hawaii Institute for Astronomy, Queen's
University Belfast, the Space Telescope Science Institute, the
South African Astronomical Observatory, and the Millennium
Institute of Astrophysics (MAS), Chile. This research has
made use of the NASA Astrophysics Data System (ADS)
Bibliographic Services and the NASA/IPAC Infrared Science
Archive (IRSA), which is funded by the National Aeronautics
and Space Administration and operated by the California
Institute of Technology. This work also made use of data
supplied by the UK Swift Science Data Centre at the
University of Leicester.

\end{acknowledgments}



%
\facilities{LCO, Gemini, ATLAS, \textit{Swift}, IRTF, MMT, BAO, Keck}

\software{astropy \citep{astropy, astropy18, astropy22}, numpy \citep{numpy}, \texttt{emcee} \citep{emcee}, HEASoft \citep{heasoft}, FLOYDS pipeline \citep{Valenti_2014}, \texttt{LCOGTSNPIPE} \citep{Valenti_2016}, matplotlib \citep{matplotlib}, Light Curve Fitting \citep{Hosseinzadeh_2020_lc, Hosseinzadeh_2023}, scipy \citep{scipy}, corner \citep{corner}, \texttt{Spextools}, \citep{Cushing_2004}.}


\appendix

\section{Tables}

\begin{table}
    \centering
    \begin{tabular}{l|c|c|l}
    \hline
         Spectrograph&Date & Phase\,/\,d  &Range\,/\AA \\
         \hline
         MMIRS & 2025-06-13& 1.2&9,800\,--\,23,500 \\
         GMOS-N& 2025-06-13 &1.5 &3,700\,--\,7,700 \\
         FLOYDS&2025-06-13  & 1.6&3,400\,--\,10,000\\
         MMIRS & 2025-06-14 & 2.2&9,800\,--\,23,500\\
         FLOYDS& 2025-06-14 &2.5 &3,400\,--\,10,000\\
         MAROON-X& 2025-06-14&2.6 &4,900\,--\,9,200 \\
         GMOS-N& 2025-06-14 &2.6 &3,700\,--\,7,700\\
         GMOS-N&2025-06-15  & 3.6&3,700\,--\,7,700\\
         MMIRS & 2025-06-16 & 4.2&9,800\,--\,23,500\\
         MAROON-X&2025-06-15  &4.2 &4,900\,--\,9,200\\
         GMOS-N& 2025-06-16 &4.6 &3,700\,--\,7,700\\
         Binospec& 2025-06-17&5.4 &4,100\,--\,9,100 \\
         Binospec& 2025-06-18& 6.2&4,100\,--\,9,100 \\
         FLOYDS & 2025-06-21& 9.7& 3,400\,--\,10,000\\
         IRTF & 2025-06-24 & 12.7&8,000\,--\,24,000\\
         GMOS-N & 2025-06-26 &14.5 &3,700\,--\,7,700\\
         FLOYDS & 2025-06-26& 14.7&3,400\,--\,10,000\\
         FLOYDS & 2025-07-01& 19.7&3,400\,--\,10,000 \\
         MMIRS & 2025-07-11 & 29.2&9,800\,--\,23,500\\
         FLOYDS & 2025-07-12 & 30.6&3,400\,--\,10,000\\
         GMOS-N & 2025-07-17 & 35.5&3,700\,--\,7,700\\
         IRTF & 2025-07-22 & 40.5&8,000\,--\,24,000\\
         FLOYDS & 2025-07-25 &43.6 &3,400\,--\,10,000\\
         GMOS-N & 2025-07-30 & 48.5&3,700\,--\,7,700\\
         MOSFIRE & 2025-08-07 & 55.5&11,000\,--\,24,000\\
         IRTF & 2025-08-07 & 56.4&8,000\,--\,24,000\\
         FLOYDS & 2025-08-10 & 59.5&3,400\,--\,10,000\\
         FLOYDS & 2025-08-23 & 72.4&3,400\,--\,10,000\\
         FLOYDS & 2025-08-24 & 73.5&3,400\,--\,10,000\\
         FLOYDS & 2025-09-08 & 88.5&3,400\,--\,10,000\\
         GMOS-N & 2025-09-12 & 92.5&3,700\,--\,7,700\\
         GMOS-N & 2025-09-14 & 94.4&3,700\,--\,7,700\\
         FLOYDS & 2025-09-14 & 94.5&3,400\,--\,10,000\\
         IRTF & 2025-09-15 & 95.5&8,000\,--\,24,000\\
         FLOYDS & 2025-09-22 & 102.4&3,400\,--\,10,000\\
         GMOS-N & 2025-10-13 & 123.4&5,300\,--\,9,700\\
         \hline
    \end{tabular}
    \caption{A summary of our spectroscopic observations throughout the evolution of SN\,2025ngs. We tabulate the spectrograph used, the date that the observations were performed and also the wavelength range of each spectrum.}
    \label{tab:spec}
\end{table}

\section{Light curve and spectral analysis}

\begin{figure}[!ht]
    \centering
    \includegraphics[width=0.32\textwidth]{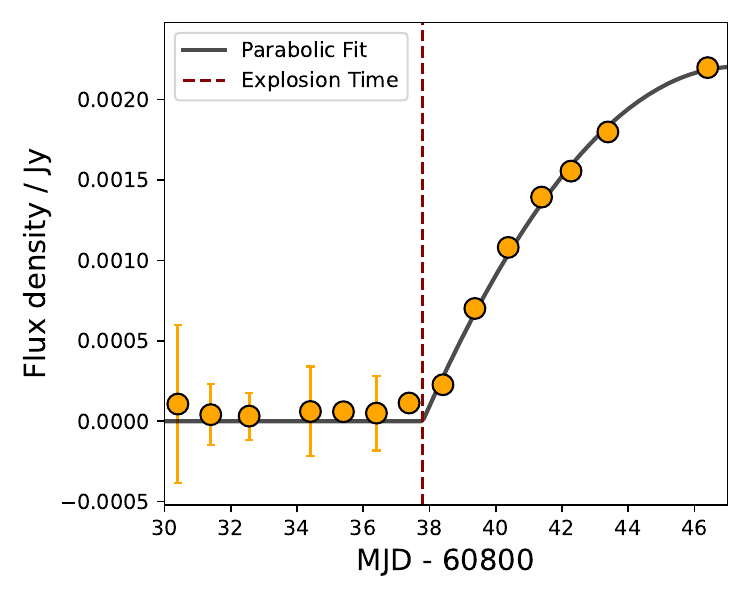}
    \includegraphics[width=0.32\textwidth]{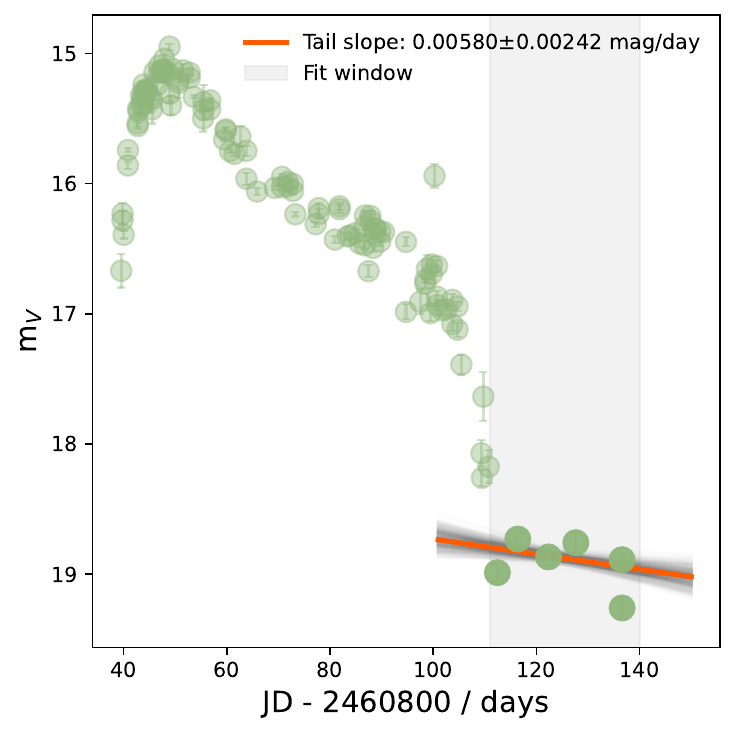}
    \includegraphics[width=0.32\textwidth]{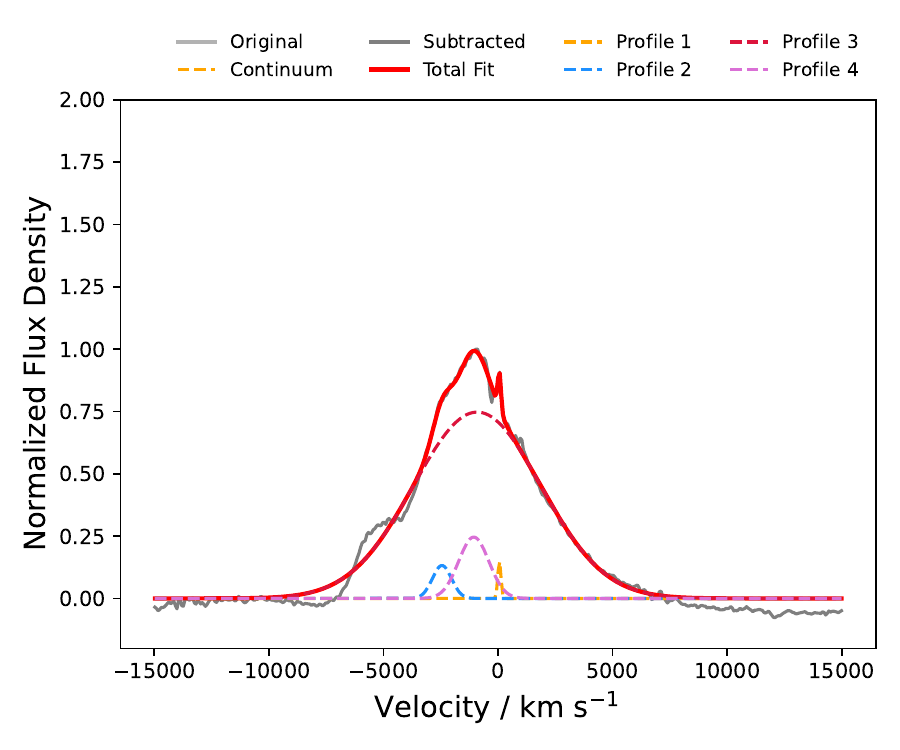}
    \caption{\textit{Left:} Our parabolic rise to the early portion of the ATLAS $o-$band light curve. \textit{Middle:} Fit to the tail portion of the late-time $V-$band light curve. \textit{Right:} Gaussian decomposition of the complex H$\alpha$ profile exhibited after the platea drop. }
    \label{fig:append}
\end{figure}

 \section{Corner Plots}



\begin{figure*}[!ht]
    \centering
    \includegraphics[width=0.99\textwidth]{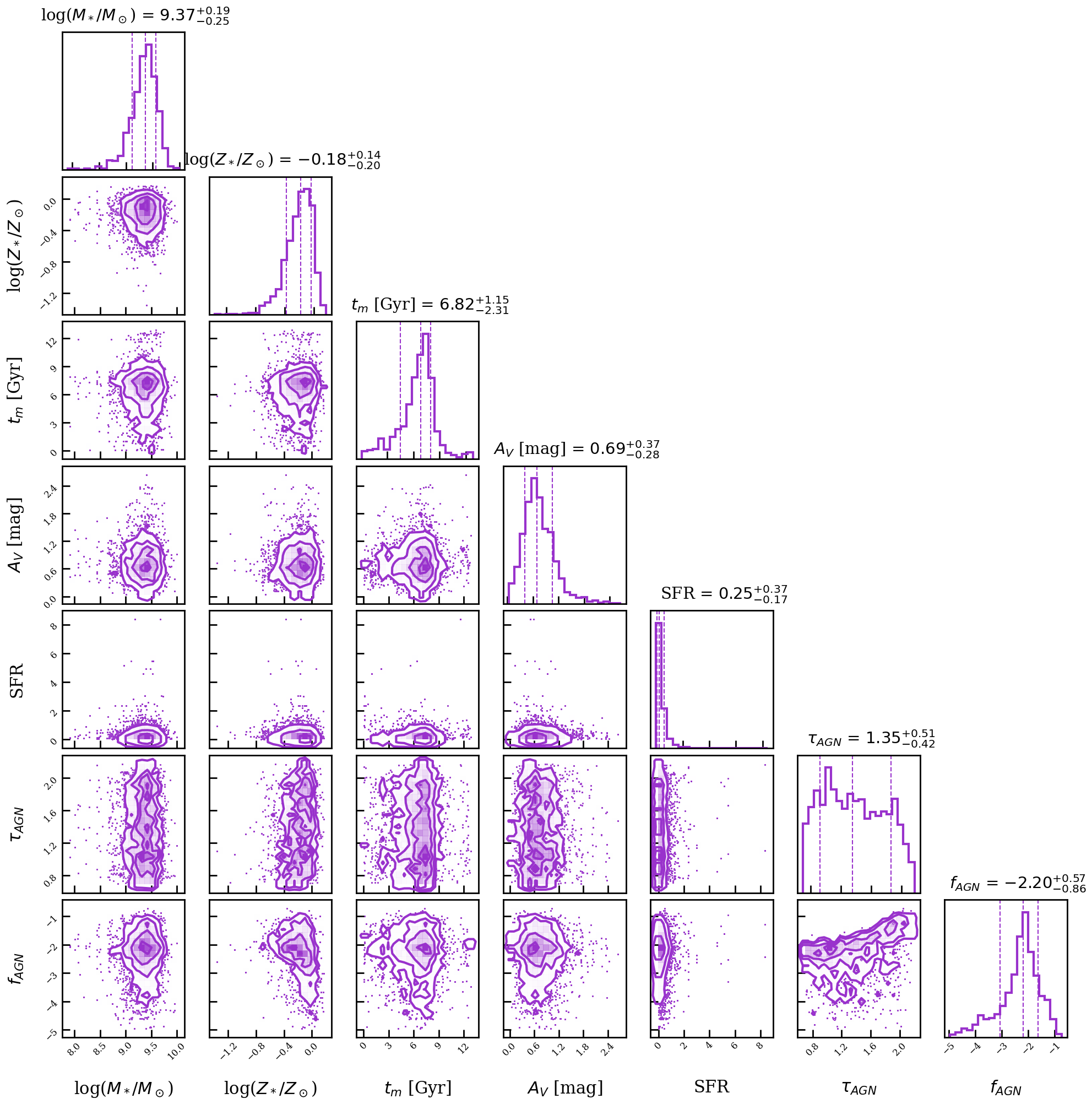}
    \caption{The corner plot for the stellar populations fit to the host of SN\,2025ngs, NGC\,5961 from \texttt{FrankenBlast}. These corner plots also show the marginal posterior distribution for each fit parameter, showing the median and spread of each posterior distribution.}
    \label{fig:blast_corner}
\end{figure*}


\bibliographystyle{aasjournalv7}
\bibliography{sample701}



\end{document}